 \documentclass[twocolumn]{aastex631}
\usepackage{amsmath}
\usepackage{breqn}

\font\manual=manfnt at 7pt \def\dbend{\hbox{\raise0.9ex\hbox{\manual\char127\hspace{0.6em}}}}

\providecommand{\e}[1]{\ensuremath{\times 10^{#1}}}

\newcounter{INTERNALionstage}

\def\gtsim{\mathrel{\hbox{\rlap{\hbox{\lower4pt\hbox{$\sim$}}}\hbox{$>$}}}}
\def\lesssim{\mathrel{\hbox{\rlap{\hbox{\lower4pt\hbox{$\sim$}}}\hbox{$<$}}}}

\def\A{{\rm\thinspace \AA}}

\def\g{{\rm\thinspace g}}

\def\K{{\rm\thinspace K}}

\def\m{{\rm\thinspace m}}

\def\s{{\rm\thinspace s}}

\def\W{{\rm\thinspace W}}

\def\h0{\mbox{{\rm H}$^0$}}
\DeclareMathAlphabet{\vib}{OML}{cmm}{m}{it}
\newcommand*{\satellite}[1]{\textit{#1}}
\newcommand*{\xmm}{\satellite{XMM-Newton}}

\renewcommand{\thefootnote}{\fnsymbol{footnote}}

\shorttitle{Non-thermal pressure
from SPT+XMM observations}
\graphicspath{{./}{figures/}}

\begin{document}

\title{
Constraints on Non-Thermal Pressure 
at galaxy cluster outskirts
from a Joint SPT and XMM-Newton Analysis}

\author[0000-0002-5222-1337]{Arnab Sarkar}
\affiliation{Kavli Institute for Astrophysics and Space Research,
Massachusetts Institute of Technology, 70 Vassar St, Cambridge, MA 02139}
\email{arnabsar@mit.edu}

\author[0000-0001-5226-8349]{Michael McDonald}
\affiliation{Kavli Institute for Astrophysics and Space Research,
Massachusetts Institute of Technology, 70 Vassar St, Cambridge, MA 02139}

\author[0000-0001-7665-5079]{Lindsey Bleem}
\affiliation{High-Energy Physics Division, Argonne National Laboratory, 9700 South Cass Avenue, Lemont, IL 60439, USA}
\affiliation{Kavli Institute for Cosmological Physics, University of Chicago, 5640 South Ellis Avenue, Chicago, IL 60637, USA}

\author[ 0000-0002-1379-4482]{Mark Bautz}
\affiliation{Kavli Institute for Astrophysics and Space Research,
Massachusetts Institute of Technology, 70 Vassar St, Cambridge, MA 02139}

\author[0000-0002-5108-6823]{Bradford A. Benson}
\affiliation{Department of Astronomy and Astrophysics, University of Chicago, 5640 South Ellis Avenue, Chicago, IL 60637, USA}
\affiliation{Kavli Institute for Cosmological Physics, University of Chicago, 5640 South Ellis Avenue, Chicago, IL 60637, USA}
\affiliation{Fermi National Accelerator Laboratory, P. O. Box 500, Batavia, IL 60510, USA}

\author[0000-0002-4469-2518]{Priyanka Chakraborty}
\affiliation{Center for Astrophysics $\vert$ Harvard \& Smithsonian, Cambridge, MA 02138, USA}

\author[0000-0002-4737-1373]{Catherine E.\ Grant}
\affiliation{Kavli Institute for Astrophysics and Space Research,
Massachusetts Institute of Technology, 70 Vassar St, Cambridge, MA 02139}

\author[0000-0003-2206-4243]{Christine Jones}
\affiliation{Center for Astrophysics $\vert$ Harvard \& Smithsonian, Cambridge, MA 02138, USA}

\author[0000-0002-9605-5588]{Florian Kéruzoré}
\affiliation{Argonne National Laboratory
9700 S. Cass Avenue, Argonne, IL 60439-4815, United States}

\author[0000-0002-3031-2326]{Eric D.\ Miller}
\affiliation{Kavli Institute for Astrophysics and Space Research,
Massachusetts Institute of Technology, 70 Vassar St, Cambridge, MA 02139}

\author[0000-0002-3984-4337]{Scott Randall}
\affiliation{Center for Astrophysics $\vert$ Harvard \& Smithsonian, Cambridge, MA 02138, USA}

\author[0000-0001-5725-0359]{Charles Romero}
\affiliation{Center for Astrophysics $\vert$ Harvard \& Smithsonian, Cambridge, MA 02138, USA}

\author[0000-0003-3521-3631]{Taweewat Somboonpanyakul}
\affiliation{Department of Physics, Faculty of Science, Chulalongkorn University, 254 Phayathai Road, Pathumwan, Bangkok 10330, Thailand}

\author[0000-0002-3886-1258]{Yuanyuan Su}
\affiliation{University of Kentucky, 505 Rose street, Lexington, KY 40506, USA}



\begin{abstract}

We present joint 
South Pole Telescope (SPT) 
and $\xmm$ observations 
of 8 massive galaxy clusters 
{ (0.8--1.7$\times$10$^{15}$
M$_{\odot}$)}
spanning a redshift range of 0.16 to 0.35. 
Employing a novel SZ+X-ray 
fitting technique,
we effectively constrain 
the thermodynamic properties of
these clusters out to the virial radius. 
The resulting best-fit electron density, 
deprojected temperature, and deprojected 
pressure profiles are in
good agreement with 
previous observations of
massive clusters. 
For the majority of the cluster sample
{ (5 out of 8 clusters)}, 
the 
entropy profiles exhibit a self-similar 
behavior near the virial radius. 
We further derive hydrostatic mass, 
gas mass, and gas fraction profiles for
all clusters up to the virial radius. 
Comparing the enclosed gas fraction 
profiles with the universal gas fraction 
profile, 
we obtain
non-thermal pressure fraction 
profiles for our cluster sample 
at  $>$$R_{500}$, demonstrating a
steeper increase 
between $R_{500}$ and $R_{200}$
that is 
consistent with the 
hydrodynamical simulations. 
Our analysis yields non-thermal pressure 
fraction ranges of 8--28\% 
{ (median: 
15 $\pm$ 11\%)} at $R_{500}$ and 
21--35\% {(median: 27
$\pm$ 12\%)} at 
$R_{200}$.
Notably, weak-lensing mass measurements are available for only four clusters in our sample, and our recovered total cluster masses, after accounting for non-thermal pressure, are consistent with these measurements.
\end{abstract}




\section{Introduction}\label{sec:intro}
Galaxy clusters, the largest gravitationally
bound structures in the universe, offer a
unique window into the complex interplay
between dark matter, galaxies, and hot
intracluster gas.
Accurate estimation of cluster
masses plays a crucial role
in understanding the properties
and dynamics of galaxy clusters
\citep[for review,][]{2019SSRv..215...25P}. 
Traditionally, cluster mass 
estimates have been derived
from X-ray and 
Sunyaev-Zel'dovich (SZ) 
observations under the assumption
that the intracluster medium (ICM) 
is in hydrostatic equilibrium with
the gravitational potential of 
the cluster
\citep[e.g.,][]{Vikhlinin_2006,2019A&A...621A..39E,2021MNRAS.501.3767S}. 
However, numerous studies have
revealed inconsistencies between
cluster 
mass estimates obtained through
hydrostatic equilibrium and those
derived from gravitational
lensing measurements
\citep[e.g.,][]{2014MNRAS.443.1973V,2016MNRAS.456L..74S,2019ApJ...884...85C}.

These inconsistencies, 
commonly known as hydrostatic 
mass bias, have raised questions 
about the validity of the
hydrostatic equilibrium assumption 
and have prompted investigations 
into the sources of such discrepancies
\citep{2021MNRAS.506.2533B}. 
It has been established that 
non-thermal pressure support in ICM, 
arising from processes such as 
turbulence, bulk motions, 
and magnetic fields, can 
significantly impact the 
estimation of cluster masses
\citep[e.g.,][]{2007ApJ...655...98N,2010ApJ...714..423K,2016ApJ...827..112B,2016MNRAS.455.2936S,2025arXiv250105514X}.
The contribution from the magnetic
field is negligible 
($\lesssim$1\%, \citealt{2014IJMPD..2330007B}), 
whereas the contribution from 
turbulence and bulk motion remains
largely unknown. 
Although there is a 
spectroscopic confirmation 
of a low-level non-thermal 
pressure, for the Perseus Cluster
using the {\it Hitomi} micro-calorimeter
\citep{2016Natur.535..117H}, 
this measurement was limited 
to the cluster center.
The micro-calorimeter onboard 
{\it XRISM}  
could provide 
crucial constraints 
on non-thermal pressure within 
R$_{2500}$\footnote{R$_{\Delta}$
is 
radius from cluster core
where matter density is 
$\Delta$ times the critical 
density of the Universe.}
\citep{2020arXiv200304962X}.
On the other hand, direct measurement of 
the non-thermal pressure support via 
line broadening
using current X-ray high-resolution 
grating spectra
is observationally expensive
at larger radii of clusters.
Future micro-calorimeter missions
with large field of view, like LEM,
will directly measure the 
spectral line broadening out to
cluster outskirts
\citep[e.g.,][]{2022arXiv221109827K,2023arXiv230701277T,2023arXiv230701259S}.

{ In recent years, the combination of 
X-ray and 
SZ 
observations of clusters
has emerged as a 
powerful approach to studying
these 
cosmic giants 
\citep[e.g.,][]{2018A&A...614A...7G,2019ApJ...871...50B,2021ApJ...918...43R,2015A&A...576A..12A,2017A&A...597A.110R,2020A&A...644A..93K,2023A&A...671A..28M}.}
In this letter, 
we explore the significance and potential of 
joint X-ray and SZ measurements in 
providing comprehensive insights into 
the physical processes governing galaxy clusters.
These two observables, X-rays and SZ signals,
probe different aspects of the cluster 
properties, enabling a multi-wavelength 
view of their structure and dynamics. 
The X-ray emission originates from the
thermal bremsstrahlung of the hot ICM, 
providing information about its temperature, 
density, and metallicity
\citep{2020ApJ...901...68C,2020ApJ...901...69C,2022MNRAS.516.3068S}.
On the other hand, the SZ effect, 
arising from the interaction of 
cosmic microwave background (CMB) photons
with the energetic electrons in the cluster, 
offers a measure of the thermal 
energy and pressure of the 
ICM \citep{1980ARA&A..18..537S}.

In this letter, we present new 
results from our joint X-ray ($\xmm$) and
SZ (SPT) observations of a sample of 
8 galaxy 
clusters at redshift 0.16 $<$ $z$ $<$ 0.35.  
For the first time, we probe detailed
thermodynamic
profiles of 
clusters,
including  mass 
measurements, and account for the 
effects of non-thermal processes
out to $R_{200}$
and beyond in this
redshift range.
{
Previous studies using
X-COP sample were
limited to redshift $<$ 0.1,
because of Planck's
large primary 
beam size 
($\sim$ 10$\arcmin$
). 
}
Throughout this paper,
we have adopted a cosmology of
$H_0$ = 70 km s$^{-1}$ Mpc$^{-1}$, 
$\Omega_{\Lambda}$ = 0.7, 
and $\Omega_{\rm m}$ = 0.3.
Unless otherwise stated, all 
reported error bars 
are at a 68\% confidence level.

\begin{figure*}
    \centering
    \includegraphics[width=0.9\textwidth]{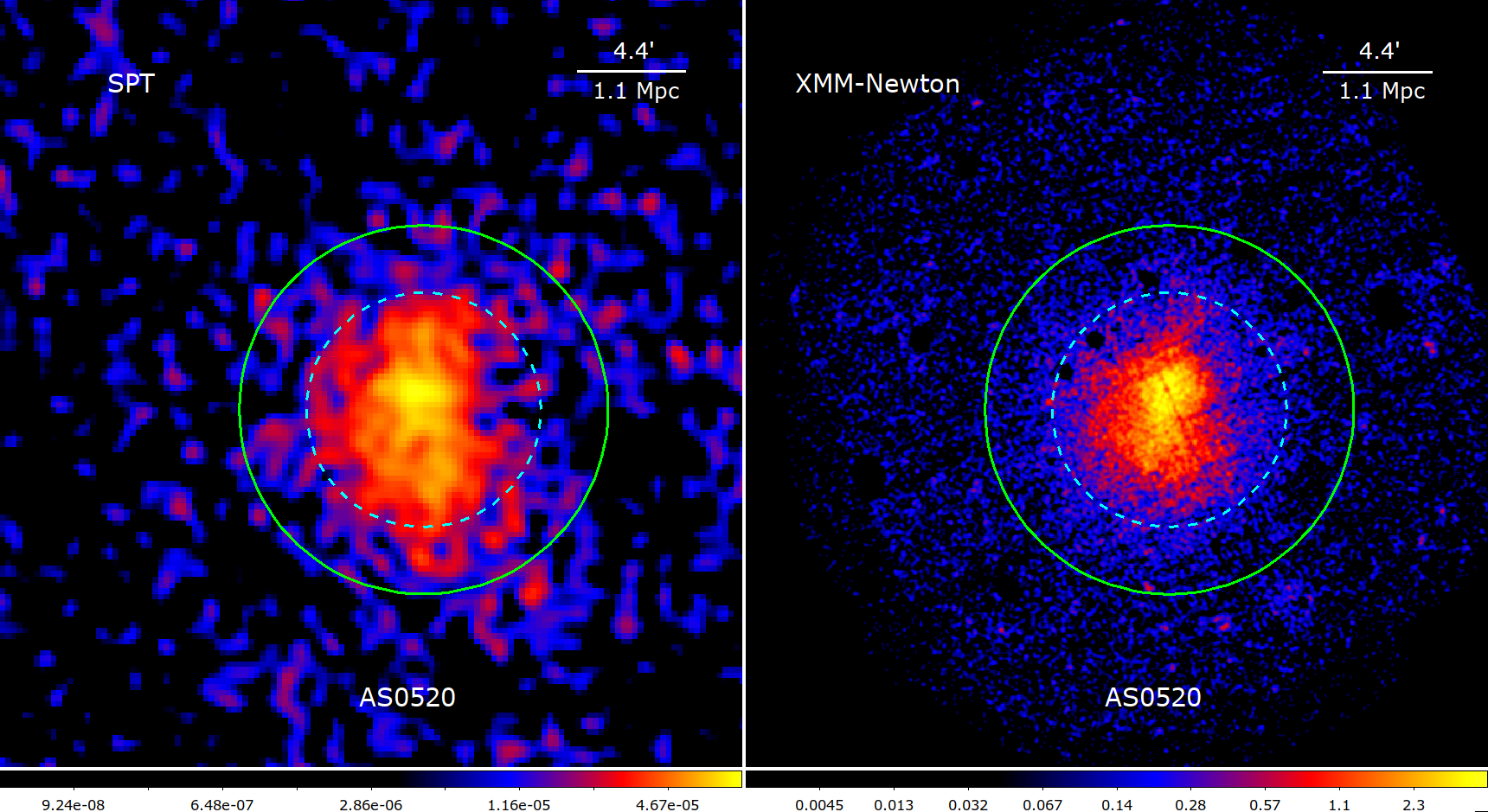}
         \begin{tabular}{cc}
        \includegraphics[width=0.5\textwidth]{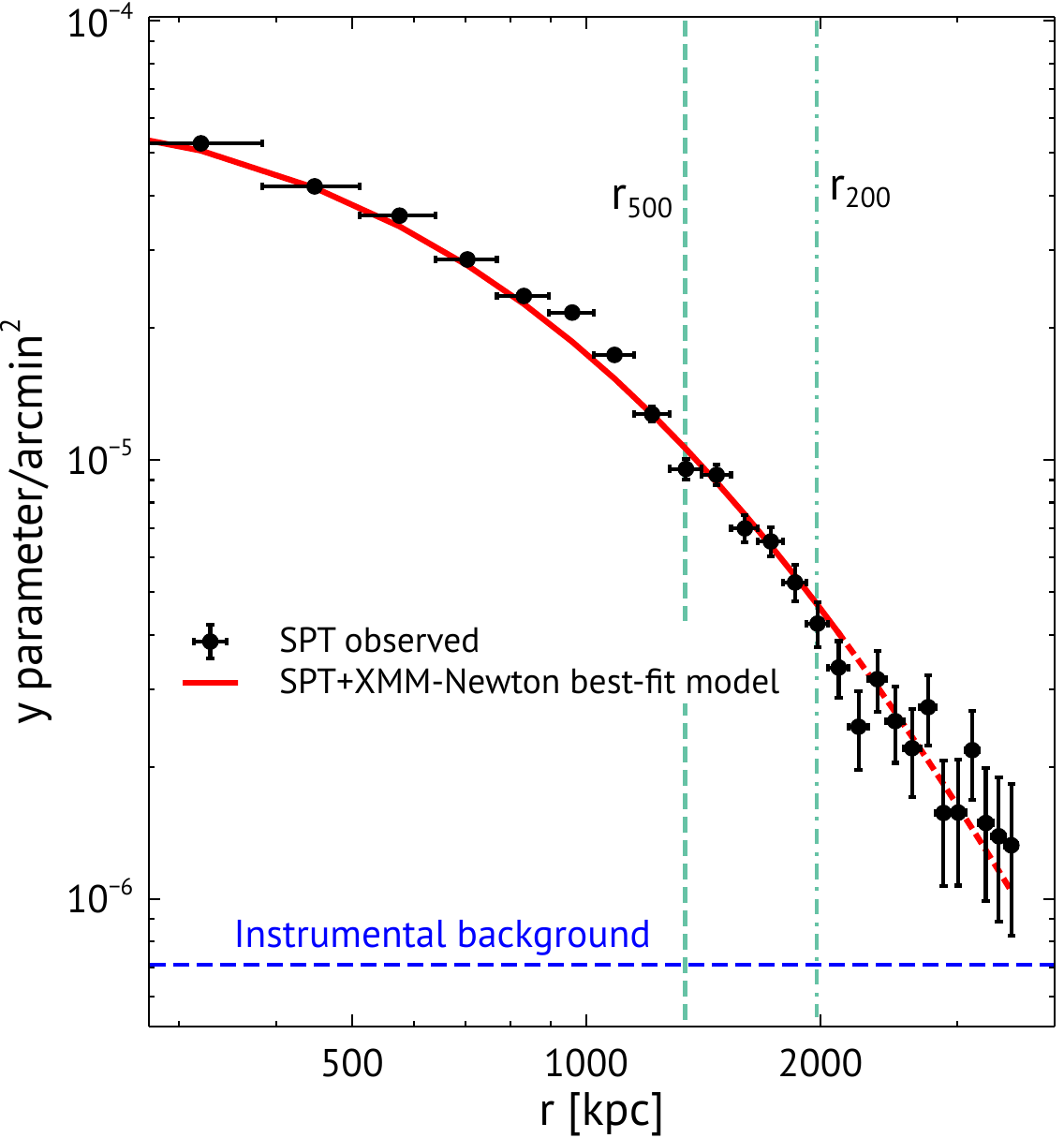}  &   \includegraphics[width=0.5\textwidth]{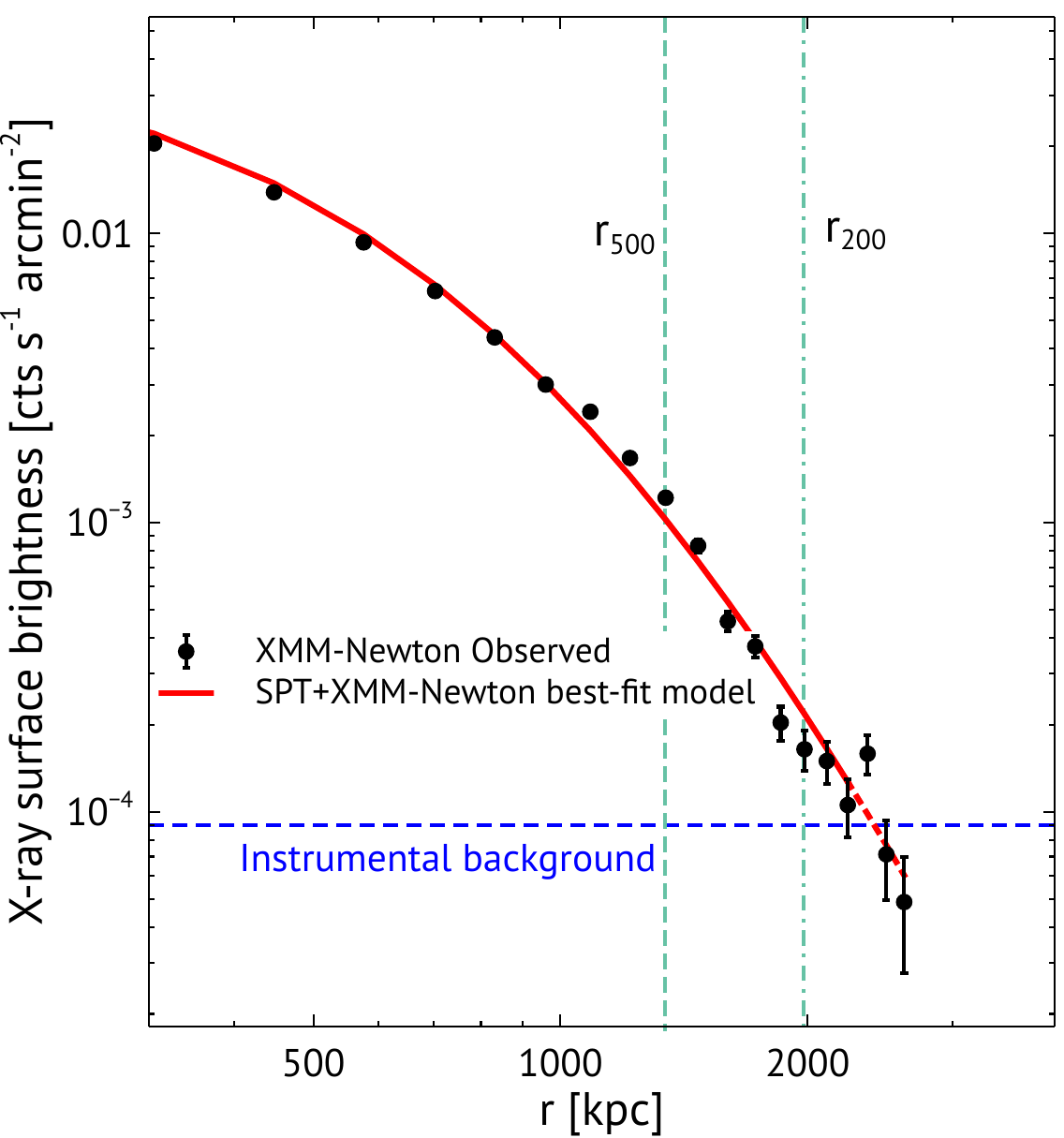}\\
    \end{tabular}
   \caption{{\it Top Left}: SPT $y$-map of 
   the galaxy cluster AS0520. 
   {\it Top Right}: 
   Quiescent particle-background subtracted 
   X-ray counts image of AS0520 in the
   0.7-1.2 keV energy band, obtained
   using $\xmm$. 
   The solid green and dashed cyan 
   circles represent R$_{200}$ and
   R$_{500}$, respectively, 
   for AS0520. 
   {\it Bottom Left}: $y$-parameter 
   profile extracted from the SPT
   SZ map of AS0520. 
   {\it Bottom Right}: 
   Similar to the bottom left figure, 
   but showing the X-ray surface 
   brightness profile of AS0520
   in 0.7-1.2 keV energy band.  
   The surface brightness profile
   is exposure corrected and background subtracted.
   { In both bottom panels, 
   the best-fit curves, as shown in solid red, are obtained by fitting only in range
   (0.15R$_{500}$--R$_{200}$). We extended the best-fit curve,
   as shown in dashed red.
   Vertical dashed and dashed-dotted
   lines represent the 
   R$_{500}$ and R$_{200}$ of AS0520, respectively. Horizontal blue dashed lines indicate respective instrumental background.}
   }
    \label{fig:AS0520_image}
\end{figure*}

\section{Sample selection and Data analysis}
{  
Our target selection criteria for 
this study include: 
(1) 
clusters with redshifts $>$ 
0.1, (2)
clusters are massive
enough to be resolved by SPT
(M$_{500}>$3$\times$
10$^{14}$M$_{\odot}$), and
(3) clusters with modest $\xmm$ 
exposures of at least 30 ks.
The chosen 
exposure time 
ensures enough photon counts
to capture X-ray 
surface brightness profiles in smaller 
radial bins with reasonable 1$\sigma$ 
uncertainty. 
For target selection, we 
focus on the 11 highest-luminosity 
REFLEX clusters at redshifts $z>$ 0.15  
studied by \citet{2010ApJ...716.1118P}. 
These clusters, observable by the SPT 
due to a range of accessible elevation 
angles, were also observed by $\xmm$. 
Out of these 11 clusters,
RXCJ0217.2-5244 and
RXCJ0528.9-3927 have $\xmm$ 
exposure times of $\sim$ 20 ks 
and 13 ks, respectively. 
Despite 
having a $\sim$58 ks $\xmm$ exposure, 
most observations of
A3888 clusters suffer from 
high background levels. 
These three 
clusters are, therefore,
not considered in this 
study. 
From the remaining clusters, 
we 
assemble a sample of 8 massive 
clusters 
within the redshift range
0.16 $<z<$ 0.35. 
} 
This joint analysis
aims
to 
investigate the gas properties of
these clusters and  
derive key physical parameters to
understand the gas dynamics better. 
The details of our cluster sample 
are listed
in Table \ref{tab:obs_log}.

\begin{table*}
    \centering
    \begin{tabular}{lcccccccc}
     Cluster ID &  RA  & Dec & $z$   & T$_{\rm X}$ & $\xmm$ exp.  & R$_{500}^{a}$ & SPT Data & Centroid shift$^{\ast}$ \\
       &   &  &   & (keV)  & (ks) & (Mpc) & Source &\\
    \hline
    \hline
    A2744 & 0$^{\rm{h}}$14$^{\rm{m}}$18.6$^{\rm{s}}$ & -30$^\circ$23$'$15.4$''$ & 0.307 & 10.1 $\pm$ 0.3$^b$ & 100 & 1.56 & P10 & $-1.54 \pm 0.01$\\
    A3404 & 6$^{\rm{h}}$45$^{\rm{m}}$30.0$^{\rm{s}}$ & -54$^\circ$13$'$42.1$''$
    & 0.164 & 8.1 $\pm$ 0.3$^b$ & 62 & 1.46 & B22 & $-2.31 \pm 0.04$\\
    AS0520 & 5$^{\rm{h}}$16$^{\rm{m}}$35.2$^{\rm{s}}$ & -54$^\circ$30$'$36.8$''$
    & 0.294 & 7.5 $\pm$ 0.3$^b$ & 67
    & 1.34 & B22 & $-1.37 \pm 0.01$\\
    AS0592 & 6$^{\rm{h}}$38$^{\rm{m}}$46.5$^{\rm{s}}$ & -53$^\circ$58$'$18.0$''$
    & 0.222 & 8.0 $\pm$ 0.4$^c$ & 48 & 1.39 & B22 & $-1.77 \pm 0.02$\\
    AS1063 & 22$^{\rm{h}}$48$^{\rm{m}}$44.9$^{\rm{s}}$ & -44$^\circ$31$'$44.4$''$
    & 0.346 & 11.1 $\pm$ 1.1$^d$ & 52 & 1.44 & P10 & $-1.76 \pm 0.03$\\
    1ES 0657--56 & 6$^{\rm{h}}$58$^{\rm{m}}$30.2$^{\rm{s}}$ & -55$^\circ$56$'$33.7$''$
    & 0.297 & 10.6 $\pm$ 0.2$^b$ & 46 & 1.57 & B22 & $-0.65 \pm 0.01$\\
    RXCJ0232.2--4420 & 2$^{\rm{h}}$32$^{\rm{m}}$18.8$^{\rm{s}}$ & -44$^\circ$20$'$51.9$''$
    & 0.284 & 7.0 $\pm$ 0.3$^b$ & 40 & 1.25 & B22 & $-2.89 \pm 0.06$\\
    RXCJ2031.8--4037 & 20$^{\rm{h}}$31$^{\rm{m}}$51.5$^{\rm{s}}$ & -40$^\circ$37$'$14.0$''$
    & 0.342 & 10.9$^e$ & 30
    & 1.2 & B22 & $-1.31 \pm 0.02$\\
    \hline
    \end{tabular}
    \caption{Galaxy cluster sample
    adopted for this study. 
    P10 and B22 are refereed to 
    \citet{2010ApJ...716.1118P} and
    \citet{2022ApJS..258...36B},
    respectively.
    Cluster redshift is taken from \citet{2010ApJ...716.1118P}.
    $^{a}$ radius measured 
    from hydrostatic mass
    profiles.
    $^{b, c, d, e}$ Gloabl X-ray temperatures
    taken from 
    \citep{2006A&A...456...55Z},
    \citep{2009AAS...21344808H},
    \citep{2008ApJS..174..117M},
    and
    \citep{2004A&A...425..367B},
    respectively.
    { $^\ast$ Centroid shift
    is measured using XMM-Newton
    data and 
    taken from
\citet{2022MNRAS.513.3013Y}.}
    }
    \label{tab:obs_log}
\end{table*}

\subsection{\it SZ: South-Pole Telescope}
The SPT is 
optimized for imaging large areas
of the CMB sky with arcminute resolution, 
and one of its primary objectives 
is the 
identification of massive clusters
via the SZ effect
\citep{2009ApJ...701...32S}. 
The galaxy clusters discussed in this
work were
observed using constant-elevation 
scans, where the telescope swept at a
constant angular velocity in azimuth 
across the field and back while stepping 
in elevation and repeating the process. 
These observations were combined 
to create single maps for each field
in different bands
\citep{2010ApJ...716.1118P,2022ApJS..258...36B}.
For this work, we utilized the
component-separated $y$-maps using a 
combination of data from the 
SPT-SZ 
and {\it Planck}
\citep{2022ApJS..258...36B}, 
which are
publicly available. 
These maps cover approximately 2500 
square degrees of the southern sky 
with a 1.25$\arcmin$ resolution and 
have been corrected for large-scale
dust emissions and point sources.
For our fiducial measurements, we 
utilized the minimum variance $y$-map 
presented in
\citet{2022ApJS..258...36B}
due to its low noise and small 
beam size, which provides optimal 
sensitivity for our analysis.

We reprojected each $y$-map 
onto the tangent plane to 
analyze individual clusters in our sample,
generating a 1024$\times$1024 pixel 
cut-out $y$-map centered on 
each cluster's position
and covering at least 20R$_{200}$. 
Figure \ref{fig:AS0520_image} 
({\it top-left})
illustrates the resulting 
$y$-map for the AS0520 galaxy
cluster as an example.
For each cluster,
we extracted the $y$-parameter radial 
profile from  
concentric annuli 
with 0.5$\arcmin$ wide
bins centered on each 
cluster core
(X-ray centroid;
\citealt{2010ApJ...716.1118P}). 
The radial range covered extends from
the cluster center to 2$R_{200}$. 
We accounted for the local background 
offset by selecting 
an area surrounding each 
cluster beyond 5$R_{200}$. 
{ Since most of the clusters in our sample are relatively hot, we 
also
accounted for relativistic corrections in the \( y \)-parameter profiles. Following \citet{2022ApJS..258...36B}, we applied a \( 5\% \) correction for clusters with \( T_{X} \leq 10 \) keV and an \( 8\% \) correction for clusters with \( T_{X} > 10 \) keV.
}
The statistical uncertainties of the
$y$-parameters were estimated by performing
random sampling of $y$-maps in
source-free regions. 
Figure \ref{fig:AS0520_image} 
({\it bottom-left}) 
shows the resulting
$y$-parameter profile for the cluster 
AS0520.

For further details regarding the 
construction, algorithms, and 
validation of the maps used 
in this analysis,
we refer the readers to 
\citet{2022ApJS..258...36B}.

\subsection{X-ray: $\xmm$} 
We used archival $\xmm$-EPIC data
spanning over two decades 
for those 8 galaxy clusters.
The data analysis for this study 
was performed using the $\xmm$
Extended Source Analysis Software 
(XMM-ESAS\footnote{\url{https://heasarc.gsfc.nasa.gov/docs/xmm/abc/}}) and related methods to 
process the EPIC 
(European Photon Imaging Camera) data. 
Initially, the event files 
were subjected to basic filtering and
calibration using XMM-ESAS tools, 
including 
{\tt epchain}, {\tt emchain}, {\tt mos-filter},
and {\tt pn-filter}. 
These tools applied the latest 
XMM-ESAS Current Calibration Files 
database to ensure accurate data 
calibration and
removed flares using the unexposed
corners of the 
instrument, using a high-energy band
from 2.5 keV to 12 keV.

The images were
created in the 0.7-1.2 keV
energy band from the filtered
event files and used to detect point
sources. Exposure maps were also
created for each detector to account
for chip gaps and mirror vignetting.
The automated point-source detection 
task {\tt cheese-bands} within XMM-ESAS 
was utilized to detect
point sources. 
All these point sources were 
excluded from further analysis.
The quiescent particle background 
(QPB) images were created from
filter-wheel closed data by employing
{\tt mos-back} and {\tt pn-back} tools
\citep{2008A&A...478..615S}.

Figure \ref{fig:AS0520_image} 
({\it top-right}) shows the resulting 
point source excluded, and QPB
subtracted count image
of AS0520. 
The local sky background were
estimated from a region surrounding
each cluster beyond R$_{200}$.
{ 
Some 
of the clusters in our sample
are undergoing mergers, 
with substructures present
within the central 
region.
We cut out the regions
obviously contaminated by 
substructures from the X-ray
image, before extracting surface
brightness profiles.
Finally, we extracted 
sky + particle background
removed 
X-ray surface brightness profiles for each 
cluster from the similar annuli bins
used for $y$-parameter profiles, as
shown in Figure \ref{fig:AS0520_image} 
({\it bottom-right}).
We adopted median technique as
proposed by 
\citet{2015MNRAS.447.2198E}
to 
extract X-ray surface brightness
profiles to
avoid
biases in 3D density profiles
due to presence of
gas clumping at cluster
outskirts.

The dynamical state of galaxy 
clusters is commonly characterized by
their proximity to virial equilibrium 
at a given time, as this can 
influence the measurement of
non-thermal pressure and, consequently,
the estimation of cluster
masses \citep[e.g.,][]{2012ApJ...751..121N,2016ApJ...827..112B}. 
In this study, we employ the centroid
shift as an
indicator of dynamical state.
The centroid shift values for our 
cluster sample were adopted from 
\citet{2022MNRAS.513.3013Y}, 
who analyzed XMM-Newton observations
of 1,308 galaxy clusters.}

\section{Joint SZ and X-ray analysis}
For each cluster in our sample, 
we utilize 
the above described 
SZ $y$-parameter and 
X-ray surface brightness
profiles to 
measure the thermodynamic profiles out
to $R_{200}$ and beyond.

\subsection{Density and temperature
profiles}
SZ observations 
directly measure the dimensionless 
Comptonization parameter, $y$, which represents
the strength of interaction between CMB 
photons and the ICM electrons. 
This Compton-$y$ parameter provides the measurement
of thermal pressure integrated along the line of
sight \citep{1972CoASP...4..173S},
\begin{equation}{\label{eq:y}}
    \centering
    y = \frac{\sigma_{T}}{m_{e}c^{2}}
    \int P\  dl,
\end{equation}
where $l$ is the distance along
the line of sight,
$\sigma_{T}$ is the Thompson 
scattering cross-section, $m_{e}$ is the
electron mass, and $c$ is the speed of light.
Thermal gas pressure, $P$, in Equation \ref{eq:y}
can be described as 
\begin{equation}{\label{eq:P}}
    P = k_{B} n_{e} T_{e},
\end{equation}
where $n_e$, $T_e$, and $k_{B}$ are the gas density,
temperature, and the Boltzmann constant, respectively.

On the other hand, 
$\xmm$ observations provide 
the X-ray surface brightness of ICM, 
which
is a function of gas density and
temperature \citep{1988xrec.book.....S},
\begin{equation}{\label{eq:sur_bri}}
    \centering
    S_{X} \approx \int n_{e}^{2}\ T_{e}^{1/2}\ dl,
\end{equation}
where the integration is along the line of sight.
One can use parametric 
expressions
of $n_e$ and $T_e$ in Equations \ref{eq:y},
\ref{eq:P}, and \ref{eq:sur_bri} and fit 
$y$-parameter and X-ray surface brightness
profiles simultaneously to put tighter
constraints on $n_e$ and $T_e$
\citep[see][]{2021ApJ...918...43R}. 
We follow this prescription for each of
the clusters in our sample. 
{ We note that temperature 
dependence in
Equation \ref{eq:sur_bri} is 
relatively mild due to the use of
narrow X-ray band.}
We adopt a gNFW density model,
as described in 
\citet{2021PhRvD.103f3514A},
\begin{equation}{\label{eq:ne}}
    n_e(r) = \frac{n_0}{(c_{500}x)^{\gamma}
    \left[1 + (c_{500}x)^{\alpha}\right]^{\frac{\beta-\gamma}{\alpha}}},
\end{equation}
where $x = r$/R$_{500}$, $c_{500}$ is the 
concentration
parameter and $\gamma$, $\alpha$, $\beta$ are the 
slopes at $r$ $\ll$ $r_s$, $r$ $\sim$ $r_s$, and
$r$ $\gg$ $r_s$, respectively, $r_s$ is a scale parameter
$r_s$ = R$_{500}/c_{500}$.

For temperature,
we adopt a universal temperature profile
as described in 
\citet{2002ApJ...579..571L},
\begin{equation}{\label{eq:temp}}
    T_e(r) = T_{0}[1+r/r_{c}]^{-\delta},
\end{equation}
where $T_{0}$, $r_{c}$,
and $\delta$ are free
parameters. 
For each cluster in our sample, 
we employ the aforementioned 
expressions for density and
temperature in Equations \ref{eq:y} 
to \ref{eq:sur_bri} and 
simultaneously fit the 
$y$-parameter and X-ray surface
brightness profiles. 
The parameters $n_0$, $r_s$, 
$\alpha$, $\beta$, $\gamma$, $T_0$, $r_c$, 
and $\delta$ in Equations \ref{eq:ne} 
and \ref{eq:temp} are allowed to vary 
freely during the fitting process.
We emphasize that
the above
temperature profile model 
accurately characterizes 
the gas temperature of the
ICM at the larger radii 
(specifically, 
$\gtrsim$ 0.15R$_{500}$).
However,
it is important to note that 
this model does not provide 
constraints on temperature 
variations within the central region
($<$ 0.15R$_{500}$).
{
We, therefore, restrict
our fitting process
within 
$0.15R_{500} < r < R_{200}$.
We note that this limitation 
does not affect the outcomes
of our study since our primary
focus is to derive non-thermal
pressure
at larger radii 
(between R$_{500}$ and R$_{200}$).
}

It's important to highlight that 
we opted not to adopt the well-known 
density and temperature profiles
by \citet{Vikhlinin_2006}.
These profiles involve 10 free parameters 
for density and 9 for temperature.
Given the context of our SZ + X-ray joint 
analysis, encompassing simultaneous 
projection and fitting with 19 free 
parameters, the computational complexity 
becomes 
formidable, and the statistical 
significance is 
limited.
Our approach with 8 free 
parameters provides sufficient 
functional freedom for the 
fitting process and better
matches the constraining power
of the data.
{ The resulting best-fit parameters
are listed in Table 
\ref{tab:best_fit_param}.
}

\begin{figure*}
    \centering
    \begin{tabular}{cc}
        \includegraphics[width=0.5\textwidth]{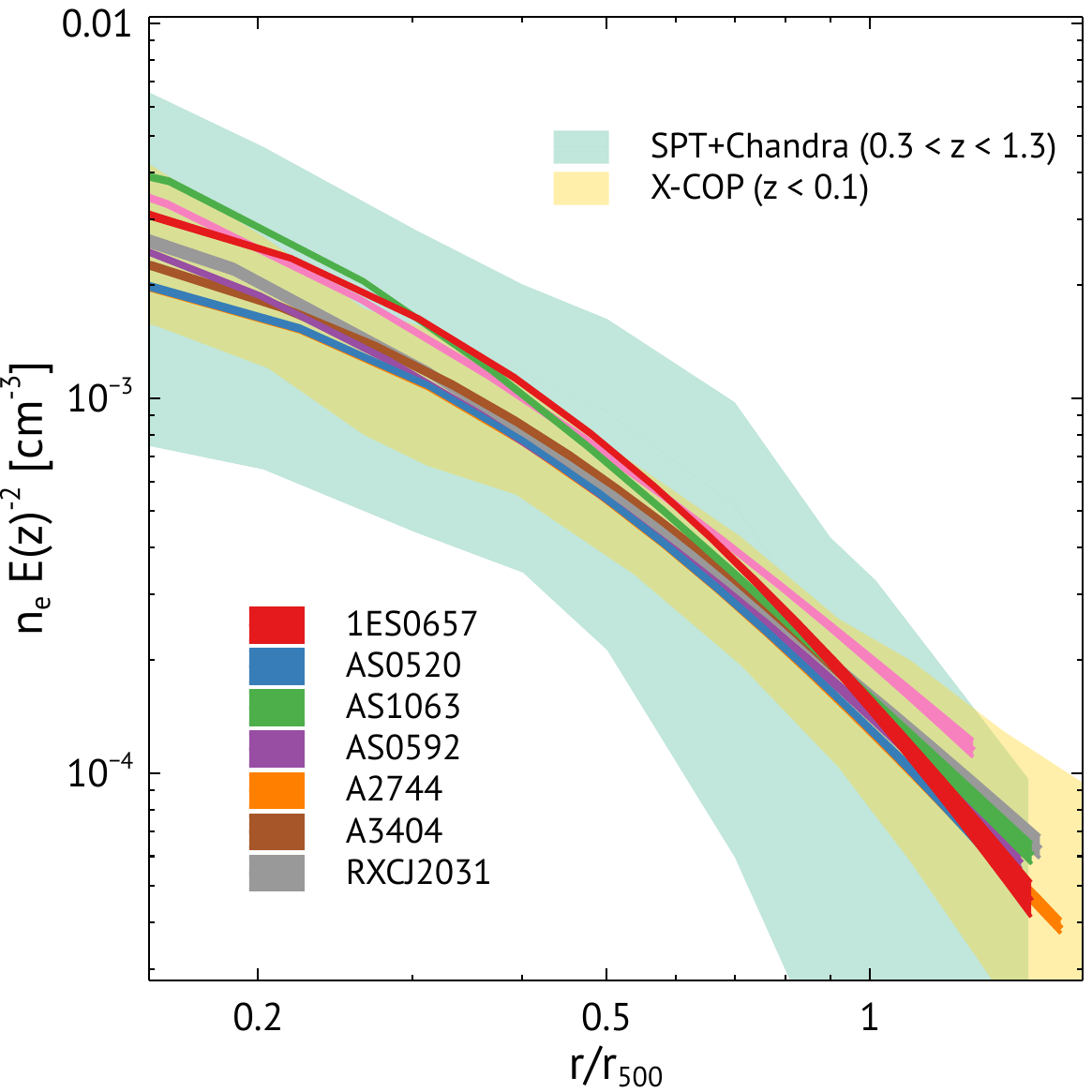} & \includegraphics[width=0.5\textwidth]{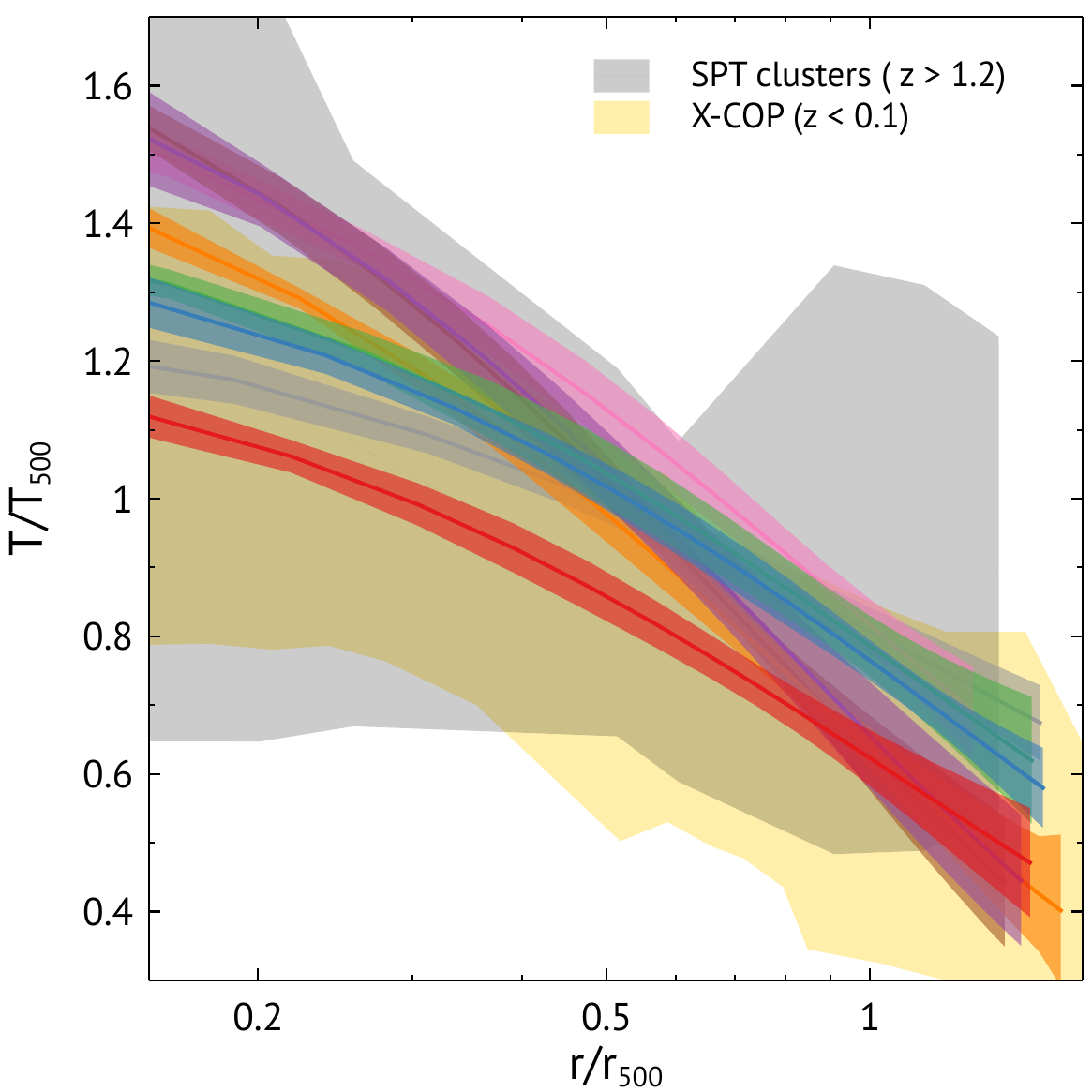}  \\
       \includegraphics[width=0.5\textwidth]{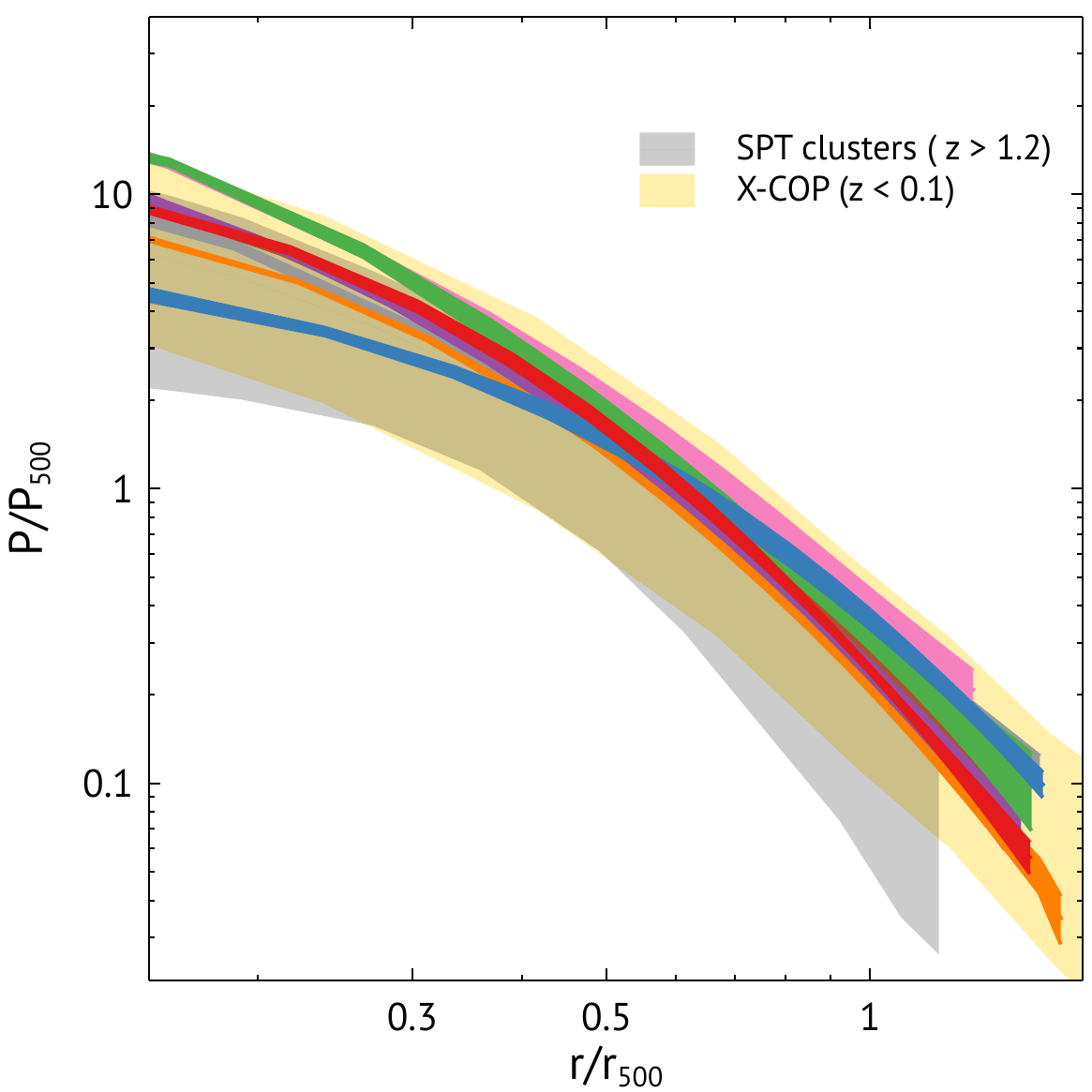}  &   \includegraphics[width=0.5\textwidth]{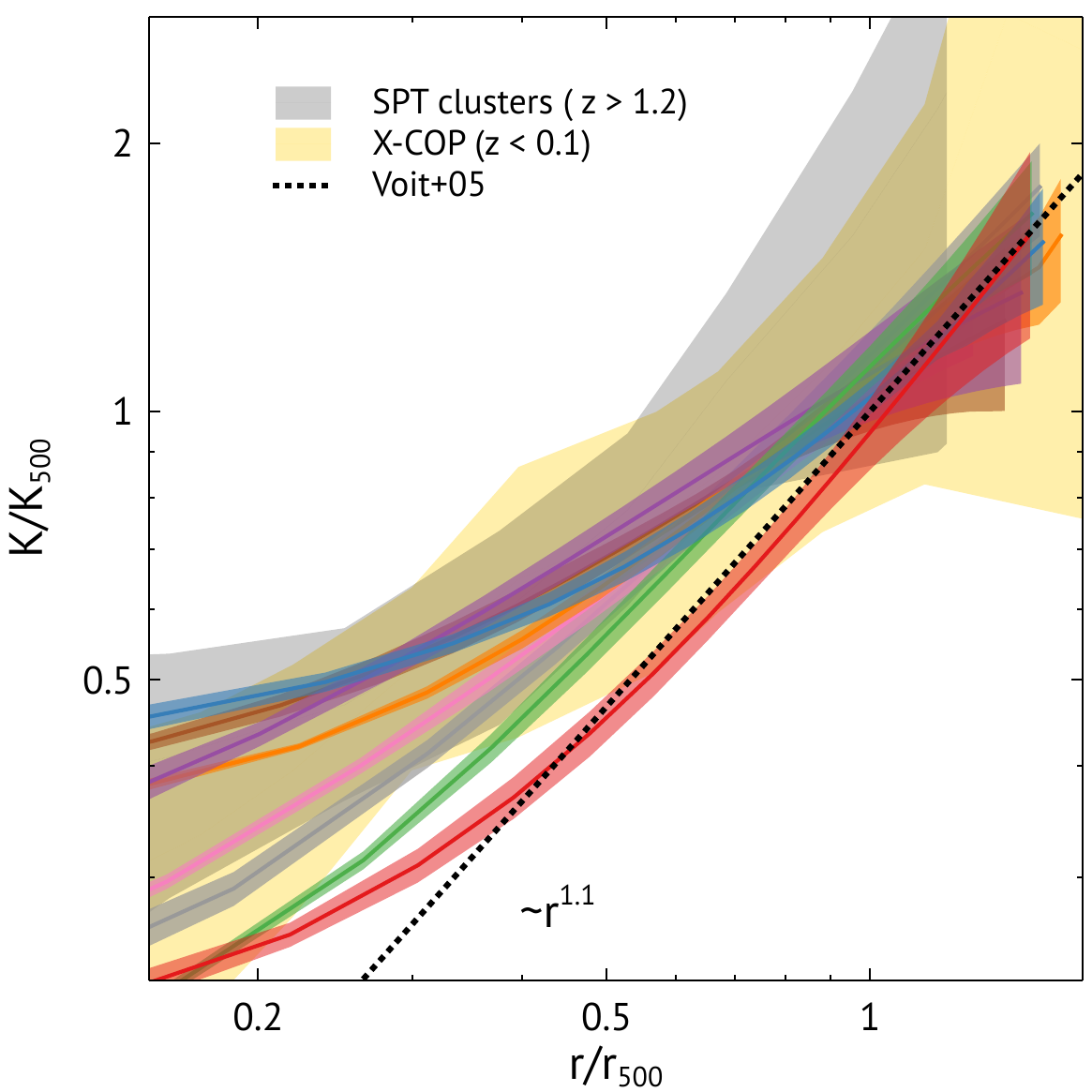}\\
    \end{tabular}
   \caption{Best-fit gas properties of galaxy clusters in our sample. {\it Top-Left}: shows the scaled
   best-fit
   electron
   density profiles.
Green shade represents density profiles
   of 67 high-redshift
   SPT clusters within ($0.3<z<1.3$)
   \citep{2021ApJ...918...43R}.
   {\it Top-Right}: scaled
   best-fit deprojected
   temperature profiles.
   {\it Bottom-Left}: scaled 
   deprojected pressure profiles.
   {\it Bottom-Right}: scaled 
   entropy profiles. Dotted line
   shows self-similar entropy profile
   \citep{2005MNRAS.364..909V}.
   In all panels,
   yellow shaded
   regions represent gas properties for X-COP sample 
   of cluster with redshift $z<0.1$
   \citep{2019A&A...621A..41G}.
   In top-right, bottom-left,
   and bottom-right panels,
   black shade regions shows 
   the gas properties of high-redshift
   ($z>$ 1.2) SPT clusters
   \citep{2021ApJ...910...14G}.
   }
    \label{fig:pressure_density}
\end{figure*}

The uncertainties are estimated using
the affine invariant
Markov Chain Monte Carlo (MCMC) sampler
implemented in the {\tt emcee} package
by
\citet{2013PASP..125..306F}.
To illustrate the results,
Figure \ref{fig:AS0520_image} 
({\it bottom} panels)
displays the best-fit 
$y$-parameter and X-ray surface brightness
profiles for cluster AS0520 as an example.
This procedure is repeated for all 8
clusters in our sample. 
The resulting best-fit scaled 
density and deprojected temperature 
 profiles are presented in 
Figure \ref{fig:pressure_density}
({\it top} panel).
Temperature profiles
are scaled by $R_{500}$ and $T_{500}$, 
where $T_{500}$ is
\begin{equation}
    T_{500} = 8.85 keV \left(\frac{M_{500}}{h^{-1}10^{15}M_{\odot}}\right)E(z)^{2/3}\left(\frac{\mu}{0.6}\right),
\end{equation}
where $M_{500}$ and $\mu$ $\sim$ 0.6
are hydrostatic
mass at $R_{500}$ 
(measured from 
hydrostatic mass profiles
shown in Figure \ref{fig:mass})
and mean molecular weight.
$E(z)$ can be expressed as 
$E(z)$ = $\sqrt{\Omega_m\left(1+z \right)^{3} + 
    \Omega_{\Lambda}}$.

Our results show
that the density 
profiles 
decline dramatically by
$\sim$ 2 dex 
from center ($>$ 0.15R$_{500}$) 
to outskirts,
consistent with the previous 
observations, such as
the X-COP sample 
\citep{2019A&A...621A..41G} 
and $z>0.3$ SPT clusters
\citep{2021ApJ...918...43R}. 
In contrast, the deprojected
temperatures experience
relatively mild variations 
across the same radial span.
Similar temperature variations 
can also
be seen in the X-COP sample 
\citep{2019A&A...621A..41G}
and in $z>1.2$ SPT clusters
\citep{2021ApJ...910...14G}.
{ 
We also derive the global temperatures of the clusters in our sample by projecting the best-fit temperature profiles onto the sky plane, as discussed in Appendix \ref{sec:tsz_txray}. We then compare our results with global temperatures obtained from spectral fits using Chandra and XMM-Newton observations. Our X-ray+SZ measurements more
closely align with the temperatures measured by XMM-Newton, as shown in Figure \ref{fig:Tx_vs_Tsz}.
}

\subsection{Pressure and entropy profiles}
If the 
ICM is unaffected by any ongoing
merger, gas pressure typically exhibits
the smoothest thermodynamic quantity
along the azimuth.
We derive 
the deprojected pressure
profiles for our cluster sample
from Equation \ref{eq:P}
by employing best-fit density and
deprojected temperature profiles.
Pressure profiles
are scaled by $R_{500}$ and $P_{500}$, 
where $P_{500}$ is
\begin{equation}
    P_{500} = 3.426 \times 10^{-3}\  {\rm keV\ cm^{-3}}\left(\frac{M_{500}}{h^{-1}10^{15}M_{\odot}}\right)E(z)^{8/3}\left(\frac{\mu}{0.6}\right).
\end{equation}
Figure \ref{fig:pressure_density} 
({\it bottom-left})
shows the deprojected pressure 
profiles of our cluster sample.
We compare our measurements with that
of the low-redshift
X-COP sample 
\citep{2019A&A...621A..41G} and 
high-redshift ($z>1.2$) SPT clusters
\citep{2021ApJ...910...14G},
as shown in Figure 
\ref{fig:pressure_density}. 
Our results are consistent with
those previous studies, except for
AS1063 and RXCJ2032 in the central
region.

Entropy, a critical parameter that 
depends on
the gas temperature ($kT$) and
electron density ($n_e$) as 
$K = kT/n_e^{2/3}$, 
plays a crucial role in 
understanding the 
thermal history of ICM. 
By examining the entropy as a 
function of radius,
we can trace the thermal evolution of the 
ICM plasma, which is subject to various 
processes such as cooling/heating, mixing, 
and convection
\citep[see][for review]{2019SSRv..215....7W}.

Numerical simulations focusing on
gravity-only structure formation 
\citep{2005MNRAS.364..909V}
have shown entropy increases radially from
the cluster center, 
following a power law. 
These simulations provide a baseline 
entropy profile prediction when clusters 
are scaled by the self-similar entropy $K_{500}$, 
resulting in the
expression 
$K(r)/K_{500}$ $\approx$ $(r/R_{500})^{1.1}$.
$K_{500}$ can be expressed as,
\begin{equation}
    K_{500} = 1667  
    \left(\frac{M_{500}}{h_{70}^{-1}10^{15}M_{\odot}}\right)^{2/3}
    E(z)^{-2/3}\ {\rm keV\ cm}^{2}
\end{equation}

Deviation from this baseline entropy
profile indicates the 
presence of non-gravitational physics
\citep[see][for review]{Nagai_2011,10.1093/pasj/63.sp3.S1019}.
In this study, we derived the entropy 
profiles of our cluster sample extended 
to $R_{200}$ and beyond, 
as shown in Figure \ref{fig:pressure_density}
({\it bottom-right}).
We find that the majority (5 out of 8) 
of the clusters
exhibit a good agreement with the baseline 
entropy profile between $R_{500}$ and $R_{200}$,
suggesting that gravitational physics predominantly 
governs the thermal evolution in this radial range. 
However, when examining individual clusters, 
we note that
both A2744 and AS0592 display mild
entropy flattening at $R_{200}$, 
although their upper limits of 1$\sigma$ errorbar
remain consistent with the baseline
profile.
In contrast, A3404 exhibits 
a significant entropy flattening that 
deviates
from the baseline profile beyond the uncertainties.
{ Our measured density profiles
are free from potential clumpy
gas at the outskirts
since we derive it
from median surface brightness
profiles 
\citep{2019A&A...621A..41G}.
This implies
thermal non-equilibrium between electrons
and ions and/or 
large scale inhomogeneities
of cluster gas distribution
may contributing to the
entropy flattening near the 
virial radius.
}

Furthermore, the central 
regions
($<$ R$_{500}$) of our cluster sample show
higher entropy values compared 
to the baseline profile.
This discrepancy 
partly suggests that 
non-gravitational physics, specifically
related to cooling, AGN feedback, 
and merging
events, 
play a substantial role in
increasing the
entropy \citep{2022ApJ...935L..23S,2023ApJ...944..132S,2023MNRAS.tmp.1967A}. 
Our findings are consistent 
with the previous 
studies 
\citep[e.g.,][]{2010A&A...511A..85P,2009ApJS..182...12C}, 
indicating that central
heating and mixing mechanisms significantly 
influence the gas properties of galaxy clusters.
We also 
find our results are consistent with
that of X-COP
and $z>1.2$ SPT clusters 
for the
entire radial range considered, 
as seen in Figure
\ref{fig:pressure_density} 
({\it bottom-right}). 
    
\subsection{Mass and $f_{\rm gas}$ profiles}

Under the assumption of the 
spherically symmetric distribution
of the 
ICM
and adherence to the equation of
state for an ideal gas, the 
combined information of gas density
and pressure enables the estimation
of the total mass of
a cluster within a given radius
\citep{1988xrec.book.....S}. 
This estimation can be
achieved through the 
hydrostatic equilibrium equation,
which accounts for the
gravitational force acting on
the gas.
We assume the ICM is in 
hydrostatic equilibrium (HSE)
within the potential well,
with the kinetic energy converted 
entirely into thermal energy
\citep{2019A&A...621A..39E}. 
In this case, hydrostatic mass
within a given radius ($r$)
can be
written as
\begin{equation}{\label{eq:hse_mass}}
    M_{\rm HSE}(<r) = -\frac{r^2}{\rho_{\rm gas}G}
    \frac{dP_{\rm th}}{dr},
\end{equation}
where $P_{\rm th}$ is the thermal electron
pressure.
$\rho_{\rm gas}$~=~1.92$\mu m_{\rm p} 
n_{\rm e}$ 
is the gas density 
\citep{2015ApJ...805..104S}. 

\begin{figure*}
    \centering
    \begin{tabular}{cc}
       \includegraphics[width=0.5\textwidth]{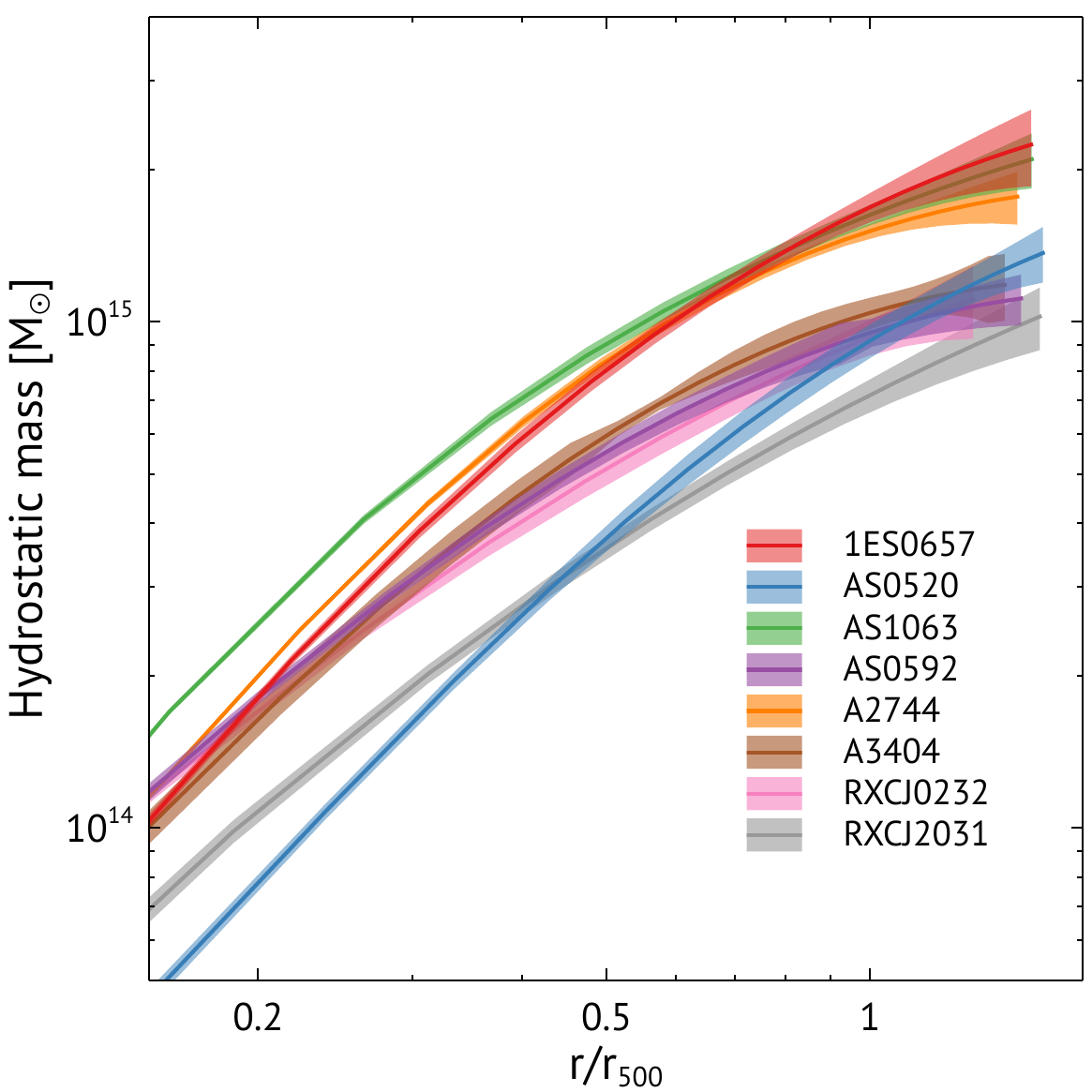}  &   \includegraphics[width=0.5\textwidth]{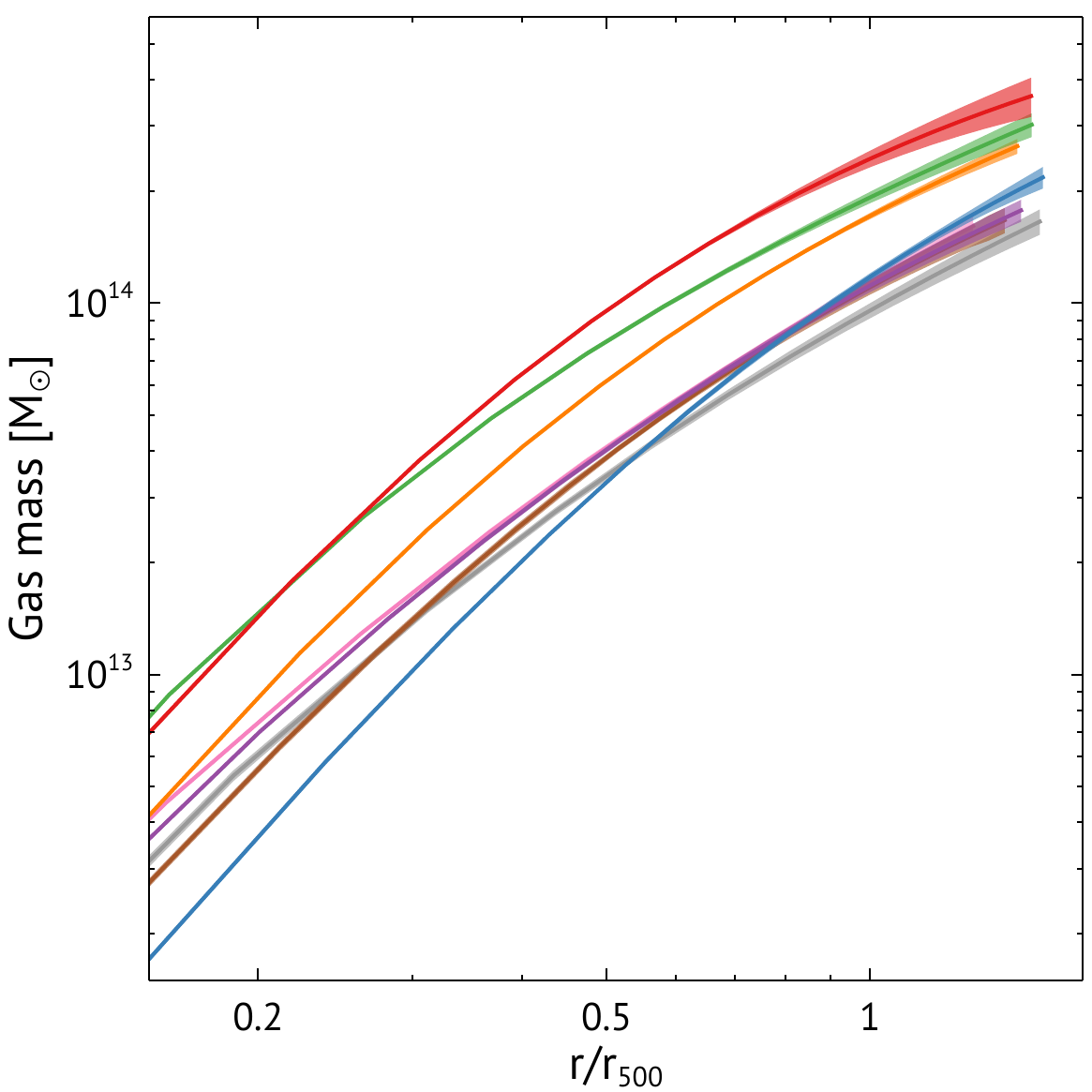}\\
       \end{tabular}
       \begin{tabular}{c}
       \includegraphics[width=0.5\textwidth]{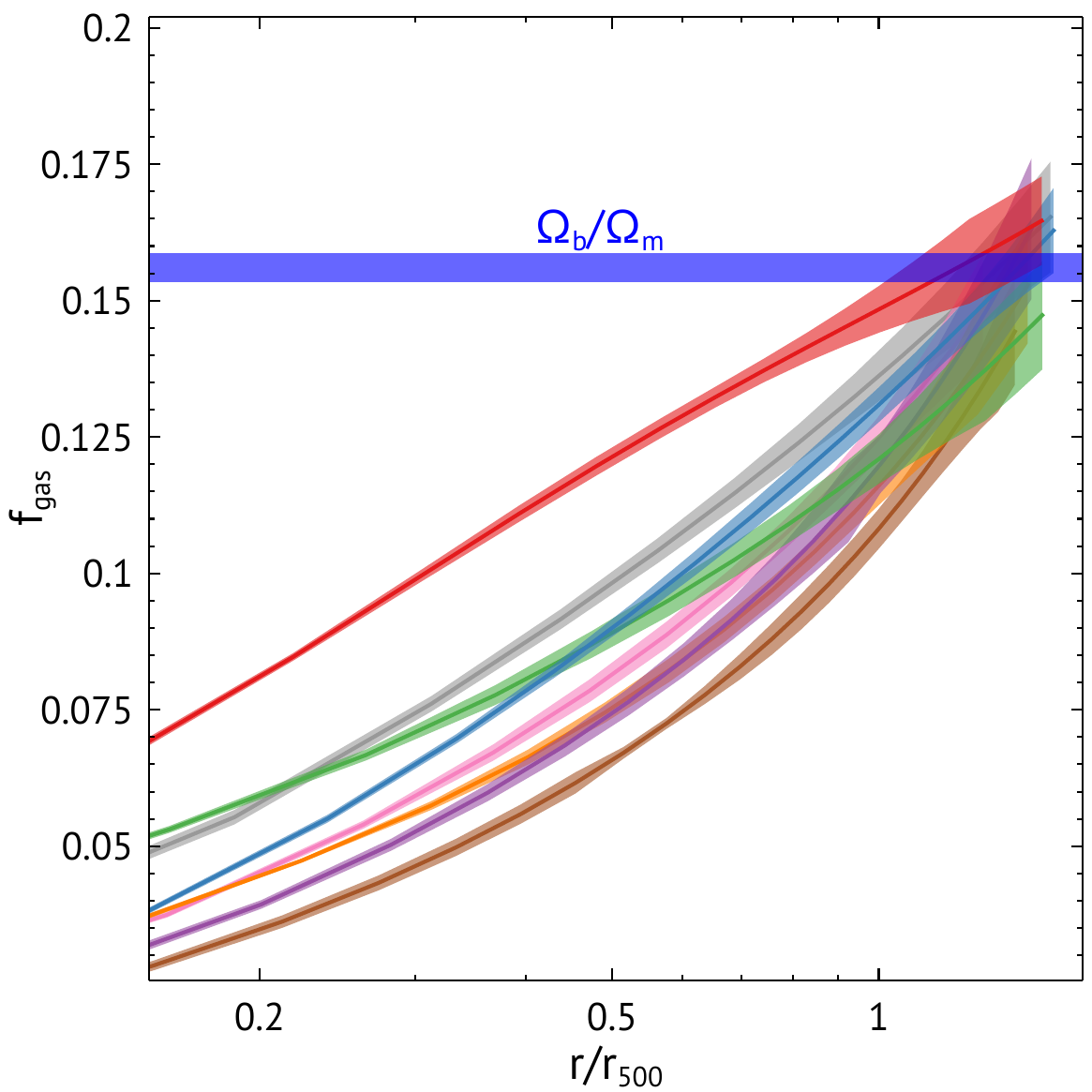} \\
    \end{tabular}
   \caption{{\it Top-Left}: hydrostatic 
   mass profiles of the 
   clusters derived using 
   best-fit pressure 
   and density profiles and assuming
   hydrostatic equilibrium. 
   {\it Top-Right}: gas mass 
   profiles
   of the sample of clusters
   using best-fit density
   profiles and assuming a
   spherical geometry.
   {\it Bottom Left}: the gas
   mass fraction
   profiles 
   of
   the sample of clusters. 
   Blue horizontal
   shaded region shows the
   {\it Planck} universal baryon 
   fraction 
    $\Omega_{\rm b}/\Omega_{\rm m}$ 
    \citep{2016A&A...594A..13P}.
    }
    \label{fig:mass}
\end{figure*}

In this study, we use
deprojected pressure and best-fit 
electron density profiles obtained from
joint SZ and X-ray fitting.
These profiles are then used
to solve Equation \ref{eq:hse_mass}
to estimate the 
hydrostatic mass for
our sample of clusters. 
Figure \ref{fig:mass} ({\it top-left})
presents the resulting
hydrostatic mass profiles 
for the cluster sample, 
spanning from the central region
to the outskirts. 
These profiles offer insights into 
the distribution of mass within
the clusters.
The high statistical quality of
the SPT and $\xmm$ data and our
novel joint fitting technique results
in overall measurement uncertainties of
$\sim$ 5\% for hydrostatic mass within 
$R_{200}$.

We also derive ICM gas mass within
a given radius ($r$) by integrating 
the best-fit gas density profile,
\begin{equation}\label{eqn:mass2}
    M_{\rm gas}(<{r}) = 4\pi\ 
    \int_{0}^{r}\ 
    \rho_{\rm gas}({r}^{\arcmin}){r}^
    {\arcmin 2} {dr}^{\arcmin},
    \end{equation}
assuming spherical symmetry 
for the clusters. 
Figure \ref{fig:mass} 
({\it top-right}) shows the 
gas mass profiles for all
8 clusters in our sample.
To further characterize the
gas content within the clusters, 
we compute the enclosed gas 
fraction profiles, defined as
$f_{\rm gas,HSE}(<r)$ = 
$M_{\rm gas}(< r)$/$M_{\rm HSE}(< r)$.
The gas fraction profile
quantifies the fraction
of the total mass composed
of ICM gas and provides valuable 
information regarding the
baryonic content of the clusters.

In Figure \ref{fig:mass}
({\it bottom}), 
we display the 
enclosed gas fraction profiles
of
our cluster
sample as a function of the 
scaled radius. 
These profiles allow us to explore
the variations in gas fraction
across different cluster scales
and provide insights into 
the distribution of baryonic matter
\citep{2010A&A...511A..85P}.
It is important to note that 
the radial range of each profile 
corresponds to regions where 
reliable information on 
density and pressure is available, 
ensuring robust estimates of the 
gas fraction. 
For our cluster sample, we are able
to measure hydrostatic mass,
gas mass, and hence gas fraction profiles
out to $\gtrsim$ R$_{200}$ without
requiring any extrapolation.
Throughout our analysis, we 
have accounted for statistical uncertainties. 
The typical uncertainties in 
$f_{\rm gas,HSE}$ 
are approximately $\leq$ 5\% at
$R_{500}$ and approximately $\leq$ 10\% at 
$R_{200}$, reflecting the precision of
our measurements and the 
reliability of the SZ + X-ray
joint fitting technique for such
high-redshift clusters.
We note that our SPT + $\xmm$
fitting process
may under-estimate the uncertainties
reported here
due to our choice of
simple temperature model.

\section{Non-thermal pressure support}
Quantifying the magnitude of
non-thermal pressure contribution 
is imperative for understanding
the mechanisms driving the 
virialization of the gas 
confined within the gravitational 
potential of the halo, 
as well as for 
accurately calibrating 
biases inherent in hydrostatic mass estimations
\citep{2014ApJ...792...25N}.
Despite the importance, the 
observational constraints on non-thermal
pressure at larger radii 
($>$0.5$R_{500}$) of high-redshift
clusters are still largely unknown.
We, therefore,
examine the non-thermal pressure support
in the ICM of our cluster sample.

\subsection{Universal gas fraction}
\begin{figure*}
    \centering
    \begin{tabular}{cc}
     \includegraphics[width=0.5\textwidth]{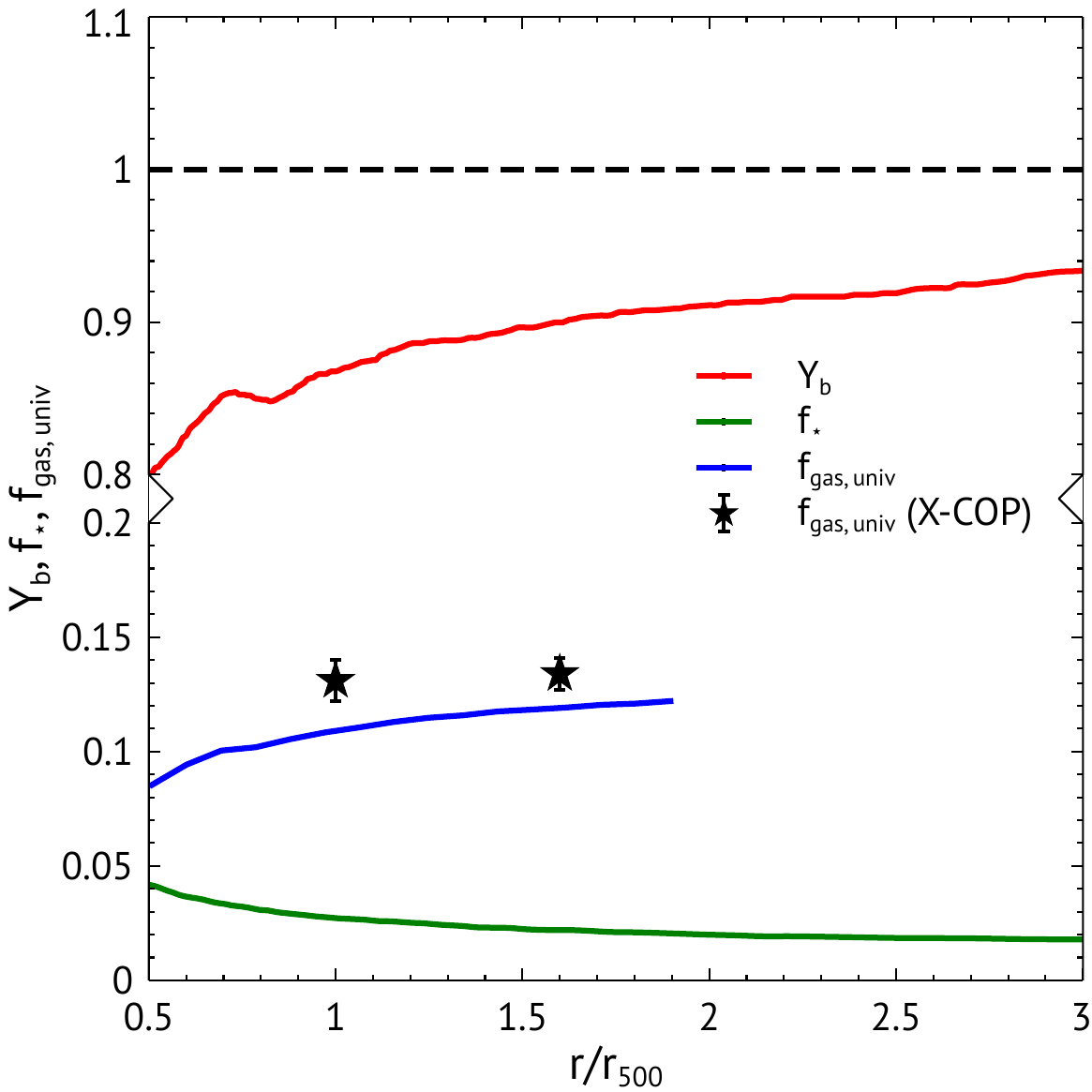} &  \includegraphics[width=0.5\textwidth]{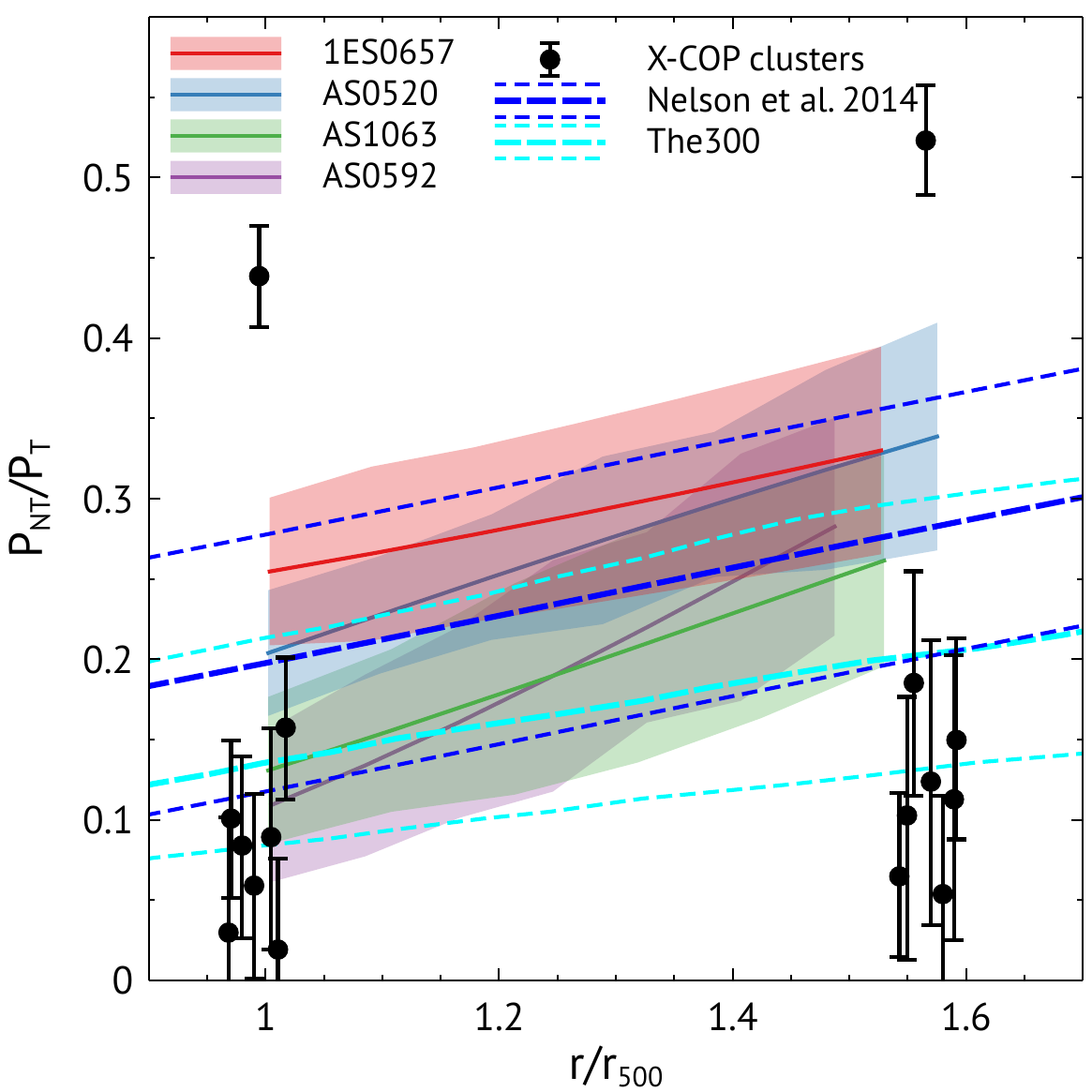} \\
    \end{tabular}
    \begin{tabular}{c}
    \includegraphics[width=0.5\textwidth]{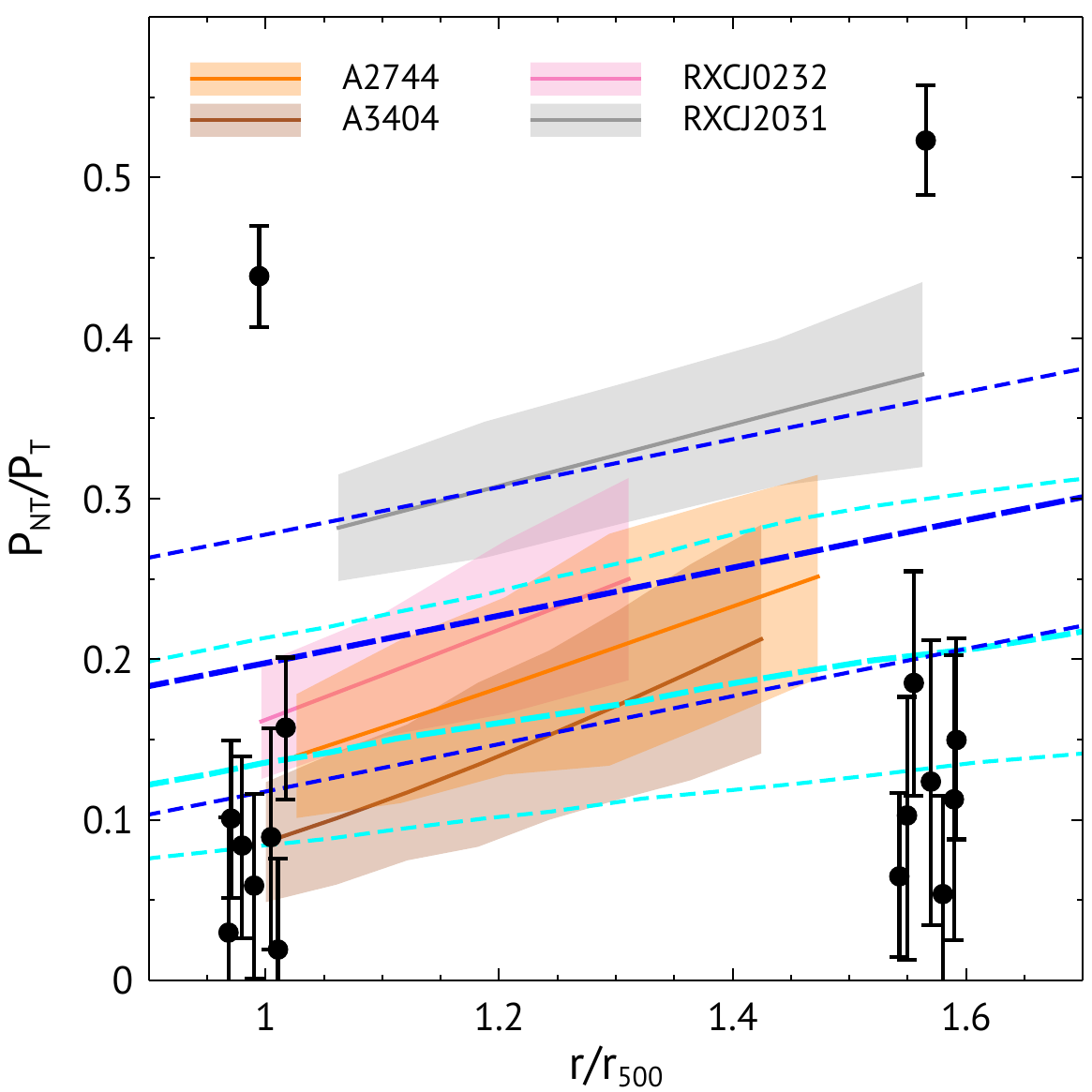} \\
    \end{tabular}
   \caption{
    {\it Top Left}: baryon depletion
    factor (red), stellar fraction (green)
    profiles as obtained from 
    \citep{2022A&A...663L...6A}. Blue curve
    shows the
    universal gas fraction. Black
    star
    represents universal gas fraction used
    for X-COP sample.
    {\it Top Right}: 
    Non-thermal pressure fraction profiles
    for 1ES0657, AS0520,
    AS1063, and AS0592. 
    The blue and cyan dashed 
    curves
    show the non-thermal pressure fractions
    predicted from the numerical simulations of
    \citet{2014ApJ...792...25N} and
    \citet{2019A&A...621A..40E}, respectively with their
    1$\sigma$ uncertainties.
    The black data
    points show the non-thermal pressure obtained
    for X-COP sample of clusters.
    {\it Bottom}:
    Similar to {\it Top Right},
    but for A2744, A3404,
    RXCJ0232, and RXCJ2031 clusters.
    }
   \label{fig:univ}
\end{figure*}

{\label{sec:univ_gas_fraction}}
The universal gas fraction within
a given radius can be written as
\begin{equation}{\label{eq:fgas}}
    f_{\rm gas,univ}(r) = Y_{b}(r)\frac{\Omega_{b}}{\Omega_{m}} - f_{\star},
\end{equation}
where $Y_{b}$ and 
$f_{\star}$
are
baryon depletion 
factor and 
fraction of baryons 
converted into
stars (or stellar 
fraction),
respectively. 
Baryon depletion factor
and stellar
fraction have been extensively studied
in the literature 
\citep[e.g.,][]{2005ApJ...625..588K,2013MNRAS.431.1487P,2015ApJ...813L..17R,2023MNRAS.tmp.1967A}.
In this present work,
we adopt the baryon depletion 
factor and stellar
fraction profiles predicted by
\citet{2022A&A...663L...6A}  for
a wide mass range, 
using
Magneticum cosmological simulation. 
Their selected sub-sample of galaxy clusters
has an average redshift of $z$ = 0.25,
consistent with the median redshift value 
$z$ = 0.29 $\pm$ 0.06 of our cluster
sample.
For this work, we specifically 
use baryon depletion
factor and stellar fraction profiles of
halos with mass $>$ 5.4 $\times$ 
10$^{14}$ h$^{-1}$ $M_{\odot}$, since our
sample consists of most massive clusters
with M$_{500}$ $>$ 5 $\times$ 10$^{14}$ 
$M_{\odot}$
(see Figure \ref{fig:mass}). 
Finally, we estimate the universal
gas fraction profile using Equation
\ref{eq:fgas}
in the 0.5 -- 2R$_{500}$ radial range.
We adopt
$\Omega_{b}$/$\Omega_{m}$ = 0.156
\citep{2016A&A...594A..13P}.
Figure \ref{fig:univ} shows the
baryon depletion factor, 
stellar fraction, and resulting 
universal gas fraction profiles 
considered in this work.
{
We use 
R$_{500}$ $< r <$ 2R$_{500}$ 
radial range of
$Y_{b}$ and $f_{\star}$ profiles
for non-thermal pressure
measurements
\citep{2016MNRAS.457.4063S}. 
}

\subsection{Non-thermal pressure fraction}
Accurate estimation of cluster 
masses from X-ray and SZ observations 
relies on the fundamental assumption
that ICM is in hydrostatic
equilibrium with the gravitational 
potential of the cluster.
However, previous studies have
highlighted discrepancies
between the mass estimates
obtained through hydrostatic 
equilibrium and those derived from 
gravitational lensing measurements
\citep{2013ApJ...767..116M,2014MNRAS.443.1973V,2014ApJ...792...25N}.
This 
hydrostatic mass bias is 
attributed to the 
non-thermal pressure support 
within the galaxy clusters. 
Unfortunately, direct
measurement of this non-thermal 
pressure using X-ray CCD/grating spectra 
of the ICM poses significant 
challenges and becomes
increasingly unfeasible at 
larger radii. 
To address this issue, we adopt a 
methodology similar to that employed by 
\citet{2014ApJ...792...25N}
to estimate the non-thermal 
pressure support for our cluster sample.

In the presence of isotropic non-thermal
pressure, the hydrostatic equilibrium
equation
can be modified as - 
\begin{equation}{\label{eq:non-therm-mass}}
    \frac{d}{dr}\left[P_{\rm th}(r) +
    P_{\rm NT}(r) \right]
     = -\rho_{\rm gas}\frac{GM_{\rm tot}(<r)}{r^2},
\end{equation}
where $P_{\rm th}$ and $P_{\rm NT}$ are 
the thermal
and non-thermal pressure, respectively,
and
$M_{\rm tot}$($<r$) is the
true total mass
of a cluster within a given radius $r$.  
The non-thermal pressure fraction,
$\alpha(r)$,
can be defined as 
the ratio between non-thermal 
pressure and total
pressure, $\alpha(r)$ = 
$\frac{P_{\rm NT}}{P_{\rm th} + P_{\rm NT}}$. 
Together with $\alpha(r)$
and Equation \ref{eq:hse_mass},
Equation \ref{eq:non-therm-mass} yields,
\begin{equation}{\label{eq:M_tot}}
    M_{\rm tot}(<r) = M_{\rm HSE}(<r)
    + \alpha(r)M_{\rm tot}(<r) - 
    \frac{P_{\rm th}r^2}{(1-\alpha)\rho_{\rm gas}G}
    \frac{d\alpha}{dr},
\end{equation}
Finally, using the formulation
of enclosed gas fraction, 
$f_{\rm gas}(<r)$ = $M_{\rm gas}(<r)$/$M_{\rm tot}(<r)$,
the equation \ref{eq:M_tot} can be simplified 
as - 
\begin{equation}{\label{eq:mass_hse}}
    M_{\rm HSE}(<r) = 
    \frac{P_{\rm th}r^2}{(1-\alpha)\rho_{\rm gas}G}
    \frac{d\alpha}{dr}
    \left[1 - \frac{f_{\rm gas,HSE}(1-\alpha)}{f_{\rm gas}} 
    \right]^{-1}
\end{equation}
Therefore, by knowing the true
gas fraction ($f_{\rm gas}$) and 
by comparing $f_{\rm gas,HSE}$,
with the universal gas fraction, one
can estimate the non-thermal pressure 
fraction, $\alpha(r)$, from Equation
\ref{eq:mass_hse}
\citep{2019A&A...621A..40E}.

We adopt a functional form of 
non-thermal pressure fraction
introduced by 
\citet{2014ApJ...792...25N},
\begin{equation}{\label{eq:alpha}}
    \alpha(r) = 1 - A\left(1 + 
    {\rm exp}\left[-  \frac{(r/R_{200})^{\gamma}}
    {B^{\gamma}}\right] \right)
\end{equation}
where $A$, $B$, $\gamma$ 
are the free parameters. 
For each cluster, we fix the
$f_{\rm gas,HSE}$
to the measured values
(Figure 
\ref{fig:mass}),
and $f_{\rm gas}$ to the 
universal gas fractions, as
derived in Section 
\ref{sec:univ_gas_fraction}.
We vary $A$, $B$, and $\gamma$ in
Equation \ref{eq:alpha} iteratively
until
Equation \ref{eq:mass_hse} 
converges to the 
measured $M_{\rm HSE}$ values.
Figure \ref{fig:univ} shows 
the
resulting non-thermal pressure
fraction profiles of 
individual clusters
between 0.75 $<$ $r/R_{500}$ 
$<$ 2$R_{500}$ radial 
range. 
Our measured non-thermal pressure 
fraction profiles
show good agreement
with that of
hydrodynamical
simulations done by 
\citet{2014ApJ...792...25N},
and 
The300 simulations shown in 
\citet{2019A&A...621A..40E}.

We further 
compare our results 
with the X-COP sample,
as illustrated in Figure 
\ref{fig:univ}. 
Specifically, at the radius $R_{500}$, 
our measurements align with those 
from X-COP within 1$\sigma$ uncertainties
for 7 out of 8 clusters, 
taking into account the spread 
observed in their sample 
(excluding A2319).
{
The non-thermal pressure 
fraction ($P_{NT}/P_{T}$)
exhibits a range of 8--23\% 
(median: 12.5 $\pm$ 6\%)
at $R_{500}$, 
with RXCJ2031 showing 
the highest value of 23 $\pm$ 2.2\%. 
\citep{2019A&A...621A..40E}
measured a median non-thermal
pressure of 
5.9$^{+2.9}_{-3.3}$\% for 
$z<$ 0.1 clusters, consistent
with our measurement.
}
Compared to the X-COP sample,
our selected SPT clusters
demonstrate a steeper increase
in $P_{NT}/P_{T}$ between the radii 
$R_{500}$ 
and $R_{200}$.
{
This discrepancy becomes
more
pronounced at $R_{200}$,
where the 
non-thermal pressure fractions 
for the SPT clusters are
relatively larger than
those observed 
in the X-COP sample
(though still overlaps 
for few clusters).
This disparity 
may be attributed
to our
adoption of
$Y_{b}$ profile
from 
\citep{2022A&A...663L...6A}.
Our measurement of $P_{NT}/P_{T}$
ranges between 21--30\% at $R_{200}$,
with a median of 25 $\pm$ 4\%.
We note that the
determination of non-thermal
pressure at the cluster
outskirts using this 
method relies significantly 
on the simulated profiles 
of $Y_{\rm b}$ and $f_{\star}$.}

\begin{figure}
    \centering
\includegraphics[width=0.48\textwidth]{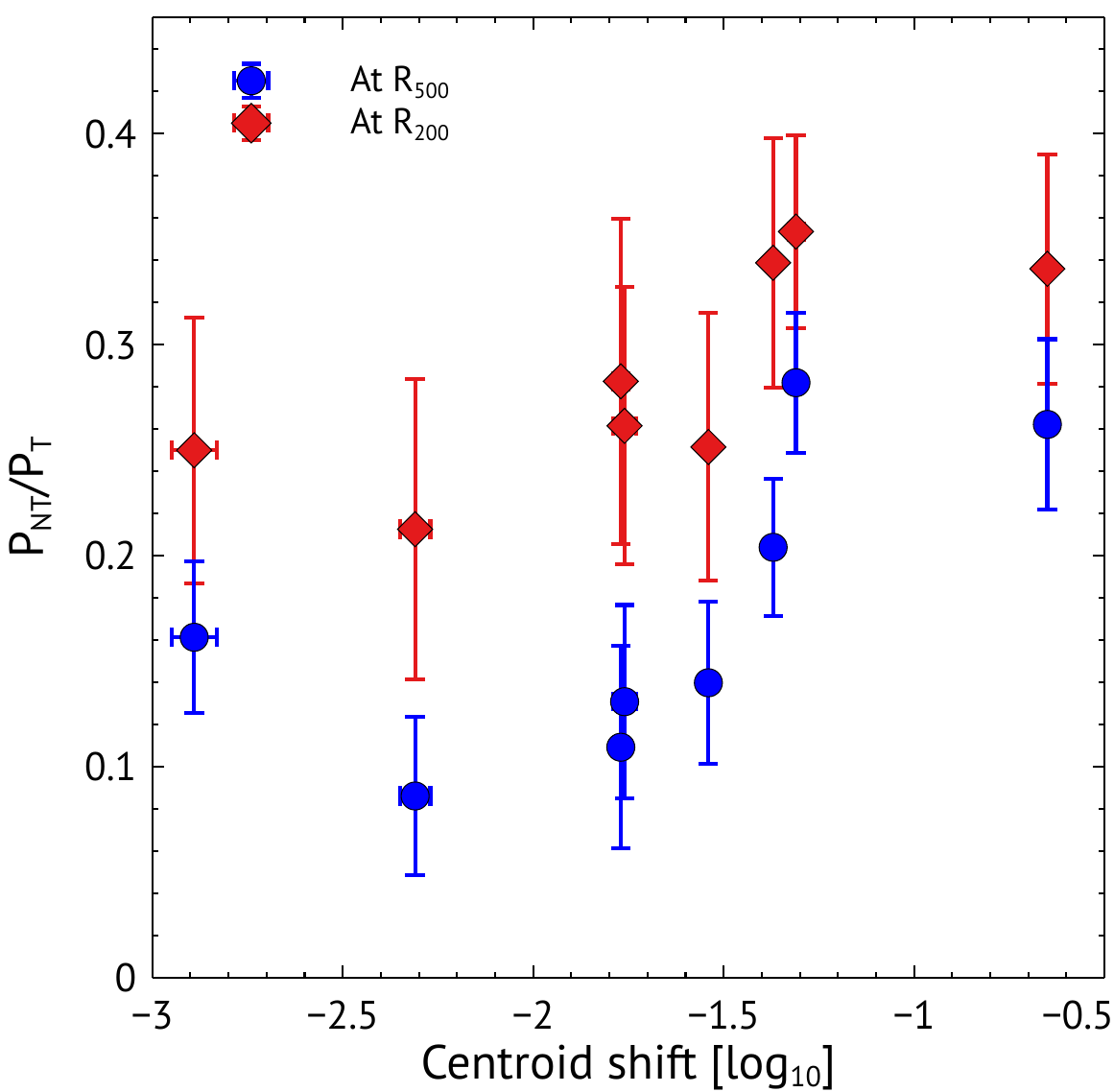} 
   \caption{Non-thermal pressure fraction of galaxy clusters in our sample as a function of centroid shift. Centroid shift values
   have been adopted from 
   \citet{2022MNRAS.513.3013Y}. 
   Higher centroid shift implies 
   cluster is more dynamically 
   active, which increases non-thermal
   pressure at cluster outskirts.
   }
    \label{fig:pnt_vs_cen}
\end{figure}

{
Next, we determine the total mass of each cluster in our sample using Equation \ref{eq:non-therm-mass} and the derived $P_{NT}/P_{T}$. The recovered total masses are listed in Table \ref{tab:physical_table}.
We then compare our recovered masses with those measured from weak-lensing (WL) observations. Since WL mass measurements are not biased by non-thermal pressure, they provide a true estimate of the cluster mass. However, not all clusters in our sample have WL mass measurements. Only four clusters—A2744 \citep{2016ApJ...817...24M}, AS0520 \citep{2009ApJ...694L.136M}, 1ES 0657-56 \citep{2006ApJ...652..937B}, and AS1063 \citep{2014MNRAS.442.1507G}—have masses measured using WL.
As shown in Table \ref{tab:physical_table}, our recovered total masses are consistent with the WL masses, demonstrating the robustness of our measurements.

The ICM of dynamically active galaxy 
clusters experiences turbulence and
bulk motion, 
increasing the non-thermal 
pressure fraction. 
Using high-resolution 
hydrodynamical simulations of 65
galaxy clusters, 
\citet{2014ApJ...792...25N} found 
a strong correlation between a
cluster's dynamical state 
and its non-thermal pressure fraction, 
with the mass accretion rate as
an indicator. 
Similarly, using centroid shift
as a dynamical state indicator,
we find that the non-thermal 
pressure fraction at R$_{500}$
increases with centroid shift,
as shown in Figure 
\ref{fig:pnt_vs_cen}.
This suggests that dynamically 
active clusters have more
turbulent ICM than relaxed ones,
consistent with simulations.
Even with large 
1$\sigma$ uncertainties,
our results hint a similar trend 
at R$_{200}$, as seen in Figure 
\ref{fig:pnt_vs_cen}. 
Future X-ray missions like NewAthena 
may better constrain non-thermal 
pressure fractions at R$_{200}$.}

{
\section{Systematic uncertainties}
We discuss different sources that
can introduce
systematic uncertainty in our 
measurements of non-thermal pressure
at cluster outskirts.
Systematic uncertainties in our
measurements can arise from 
subtracting 
XMM-Newton quiescent particle background.
\citet{2018A&A...614A...7G}
showed that 
residual soft-proton background 
remains (at $\sim$ 15\%) 
after subtracting 
QPB, if the observation
is impacted by severe soft-proton
background. 
We, therefore, only include radial
bins that are within R$_{200}$ of
the respective clusters to avoid
including bins where count rates
are dominated
by soft-proton background.
For all the clusters in our sample,
the sky-background subtracted source
count rates are a factor of
$\geq$ 2 (i.e, $\geq$ 3.5$\sigma$) 
higher than the QPB level
at R$_{200}$, as shown in Figure 
\ref{fig:AS0520_image}. 
We also adopt a systematic 
uncertainty of 15\% of the 
non X-ray background
level on the measured 
X-ray surface brightness profiles.

Another source of systematic
uncertainty may stem from our 
choices of simple electron
temperature
and density models to fit X-ray
surface brightness and SZ y-map
profiles. 
\citet{2019A&A...621A..41G} 
showed that slope of density
profiles steepen with radius
even at larger radii, but slope
of temperature profiles stay 
relatively flat at 
r $>$ 0.3R$_{500}$.
Our gNFW based density model 
does not capture any variation
in slopes of density profiles
at $r \gg r_{\rm s}$, which may
underestimate the uncertainty
levels on density and temperature
profiles at cluster outskirts.

One of major source of systematic
uncertainties in non-thermal pressure
measurements
stem from adopted
values of $Y_{b}$ and $f_{\ast}$
from 
hydrodynamical simulations. 
As seen in Figure \ref{fig:univ},
our measurements for $z>$ 0.1
clusters are systematically shifted
to larger $P_{NT}/P_{T}$ values
compared to X-COP clusters because
of different choices of 
$Y_{b}$ and $f_{\ast}$ values between
two studies.

}

\section{Conclusion and summary}

X-ray observations offer a precise 
estimation of ICM gas density,
while SZ observations provide 
accurate measurements of thermal 
pressure out to $R_{200}$ and beyond
\citep{2018A&A...614A...7G,2022ApJS..258...36B}. 
By integrating these complementary
data sources, we can potentially 
enhance our understanding of cluster 
formation and evolution. 
In this study, we have 
introduced a
joint fitting technique that combines
X-ray and SZ data to obtain precise 
constraints on ICM gas properties. 
To validate our approach,
we apply this technique to a
sample of 8 massive galaxy 
clusters (0.16 $<z<$ 0.35) 
carefully selected from
the SPT and XMM cluster catalogs.
Here, we present a summary of our 
key findings below.

\begin{itemize}
    \item We show that
    a gNFW electron density
    model \citep{2021PhRvD.103f3514A},
    together with
    a universal temperature profile
    as described in 
    \citet{2002ApJ...579..571L}, fits best
    with the observed
    X-ray surface brightness
    and $y$-parameter profiles out to
    $R_{200}$ when fitted jointly.
    The best-fit density and temperature
    profiles are in good agreement 
    with previous other studies
    \citep[e.g.,][]{2019A&A...621A..41G,2021ApJ...910...14G,2021ApJ...918...43R}.

    \item The pressure profiles
    of our cluster sample show similar
    trends, as expected for massive
    galaxy clusters 
    ($M_{500}>$3$\times$10$^{14}$ 
    M$_{\odot}$). Our measurements 
    of pressure profiles are consistent
    with that of the X-COP cluster
    sample \citep{2019A&A...621A..41G} 
    and $z>1.2$ SPT cluster
    sample \citep{2021ApJ...910...14G}.
    We find that the majority
    of the clusters
    in our sample 
    show a good agreement with the baseline
    entropy profile between $R_{500}$
    and $R_{200}$, indicating that 
    gravitational
    collapse governs the ICM 
    heating in this radial
    range.

    \item Our measurements of $M_{\rm HSE}$ 
    for the cluster sample range
    from 
    0.75--1.68$\times$10$^{15}$ M$_{\odot}$ 
    within
    $R_{500}$ and 1--2.2$\times$10$^{15}$ M$_{\odot}$ 
    within
    $R_{200}$. 
The $f_{\rm gas}$ profiles
    of our cluster sample increase 
    from the cluster centers and reach
    values between 
    0.108--0.148 at
    $R_{500}$ and 
    0.144--0.165
    at $R_{200}$.

    \item 
    We adopted $Y_{\rm b}$ and
    $f_{\star}$ profiles from
    Magneticum simulation
    \citep{2022A&A...663L...6A}
    for
    deriving non-thermal pressure.
    Our estimated non-thermal
    pressure fraction ($P_{NT}$/$P_{T}$) 
    ranges from 
    8\% to 28\% (median: 15 $\pm$ 11\%)
    at $R_{500}$ and from 21\% to 35\%
    (median: 27 $\pm$ 12\%) at
    $R_{200}$, consistent with
    the hydrodynamical 
    simulations
\citep{2014ApJ...792...25N,2019A&A...621A..40E}. Our results 
    at $R_{500}$
    are consistent with the
    X-COP measurements within 1$\sigma$ uncertainties, but
    differ significantly 
    at $R_{200}$. 
    We note that our results show a steeper
    increase in $P_{NT}$/$P_{T}$ between
    $R_{500}$ and $R_{200}$ compared to
    the X-COP sample.
    This disagreement likely stems
    from differing
    choices of 
    $Y_{\rm b}$,
    and $f_{\star}$ profiles
    in the two studies.
    { We recover the total cluster mass for each cluster in our sample, accounting for non-thermal pressure, and compare these values with masses derived from weak-lensing observations. The consistency between our recovered cluster masses and the available weak-lensing masses demonstrates the robustness of our measurements.
    }
    
\end{itemize}

\begin{acknowledgements}
We sincerely thank the anonymous
referee for their insightful 
comments and suggestions.
A.S gratefully acknowledge 
support from NASA grant     
80GSFC23CA045 and 
Smithsonian Astrophysical 
Observatory sub-award 
SV2-82023.
This work is based on 
observations obtained with 
the South Pole Telescope and
XMM-Newton.
The South Pole Telescope program
is supported by the National 
Science Foundation (NSF)
through the 
Grant No. OPP-1852617.
XMM-Newton is an ESA science 
mission with instruments and 
contributions directly 
funded by ESA member states 
and the USA (NASA).
Work at Argonne National Lab 
is supported by UChicago 
Argonne LLC, Operator of 
Argonne National Laboratory 
(Argonne). 
Argonne, a U.S. Department
of Energy Office of Science 
Laboratory, is operated 
under contract no. 
DE-AC02-06CH11357.

\end{acknowledgements}

\bibliographystyle{mnras}
\bibliography{sample631} 

\appendix
{ 
\section{Physical properties of our sample of clusters}\label{tab:physical_table}
\begin{table*}[h!]
    \centering
    \footnotesize
    \begin{tabular}{lcccccccc}
     Cluster  & M$_{\rm HSE,500}$\ \ & M$_{\rm HSE,200}$ \ \  & $f_{\rm gas,500}$\ \ & $f_{\rm gas,200}$\ \  & $\frac{P_{\rm NT}}{P_{\rm T}}_{500}$\ \ & $\frac{P_{\rm NT}}{P_{\rm T}}_{200}$ & M$_{\rm recovered,200}$\ \ & M$_{\rm WL}$\\
       & ($10^{15} M_{\odot}$) & ($10^{15} M_{\odot}$) &  & & ($\%$) & ($\%$) & ($10^{15} M_{\odot}$) & ($10^{15} M_{\odot}$)\\
    \hline
    \hline
    A2744 & 1.45 $\pm$ 0.06 & 1.76 $\pm$ 0.18 & 0.111 $\pm$ 0.005 & 0.153 $\pm$ 0.011 & 13.9 $\pm$ 3.8 & 25.1 $\pm$ 6.3 & 2.22 $\pm$ 0.63 & 2.1 $\pm$ 0.4$^a$\\
    
    A3404 & 1.04 $\pm$ 0.07 & 1.21 $\pm$ 0.16 & 0.108 $\pm$ 0.006 & 0.144 $\pm$ 0.010 & 8.6 $\pm$ 3.7 & 21.2 $\pm$ 7.1 & 1.31 $\pm$ 0.47 & --\\
    
    AS0520 & 0.92 $\pm$ 0.08 & 1.37 $\pm$ 0.15 & 0.131 $\pm$ 0.005 & 0.163 $\pm$ 0.009 & 20.4 $\pm$ 3.3 & 33.8 $\pm$ 5.9 & 2.13 $\pm$ 0.44 & 1.7$^{b}$\\
    
    AS0592 & 0.96 $\pm$ 0.06 & 1.12 $\pm$ 0.13 & 0.120 $\pm$ 0.007 & 0.161 $\pm$ 0.012 & 10.9 $\pm$ 4.7 & 28.3 $\pm$ 7.7 & 1.51 $\pm$ 0.45 & --\\
    
    AS1063 & 1.63 $\pm$ 0.07 & 2.13 $\pm$ 0.26 & 0.121 $\pm$ 0.006 & 0.147 $\pm$ 0.010 & 13.1 $\pm$ 4.6 & 26.1 $\pm$ 6.6 & 2.63 $\pm$ 0.65 & 2.3 $\pm$ 0.4$^{d}$\\
    
    1ES 0657--56 & 1.68 $\pm$ 0.11 & 2.23 $\pm$ 0.38 & 0.148 $\pm$ 0.007 & 0.164 $\pm$ 0.012 & 25.4 $\pm$ 5.0 & 34.2 $\pm$ 8.5 & 3.15 $\pm$ 0.71 & 3.1 $\pm$ 0.5$^{c}$\\
    
    RXCJ0232.2--4420 & 0.94 $\pm$ 0.06 & 1.11 $\pm$ 0.18 & 0.124 $\pm$ 0.005 & 0.149 $\pm$ 0.009 & 16.1 $\pm$ 3.6 & 24.9 $\pm$ 6.3 & 1.33 $\pm$ 0.43 & --\\
    
    RXCJ2031.8--4037 & 0.75 $\pm$ 0.07 & 1.02 $\pm$ 0.14 & 0.139 $\pm$ 0.005 & 0.165 $\pm$ 0.009 & 28.2 $\pm$ 3.3 & 35.3 $\pm$ 5.5 & 1.74 $\pm$ 0.47 & --\\
    \hline
    \end{tabular}
    \caption{Galaxy cluster sample
    adopted for this study. 
    $\frac{P_{\rm NT}}{P_{\rm T}}_{500}$ and $\frac{P_{\rm NT}}{P_{\rm T}}_{200}$ are non-thermal pressure fractions at 
    R$_{500}$ and R$_{200}$, respectively.
    M$_{\rm recovered,200}$
    is the recovered
    cluster total mass after 
    accounting for non-thermal 
    pressure. 
    M$_{\rm WL}$ represents cluster 
    mass measured from weak-lensing (WL)
    observations within R$_{200}$
    adopted from:
    $^{a}$ \citet{2016ApJ...817...24M},
    $^{b}$
    \citet{2009ApJ...694L.136M},
    $^{c}$ \citet{2006ApJ...652..937B},
    $^{d}$ \citet{2014MNRAS.442.1507G}.
    }
\end{table*}

\section{best-fit 
parameters from SZ y-map + X-ray surface brightness fitting}\label{tab:best_fit_param}
\begin{table*}[h!]
    \centering
    \footnotesize
    \begin{tabular}{lcccccccc}
     Cluster & log$(\frac{n_0}{ E(z)^{2}})$ & $\alpha$ & $\beta$ & $\gamma$ & $r_s$ & $T_0$ & $r_c$ & $\delta$ \\
       &   &  &   &  \ \ & (Mpc) & (keV) & (Mpc)  &\\
    \hline
    \hline
    A2744 & $-2.53$ & 2.0 & 2.45 & 0.13 & 0.54 & 14.9 & 4.5 & 2.55\\
    
    A3404 & $-2.46$ & 1.44 & 2.8 & 0.12 & 0.60 & 16.8 & 2.8 & 2.6\\
    
    AS0520 & $-2.71$ & 1.41 & 3.3 & 0.11 & 1.0 & 11.8 & 2.6 & 1.64\\
    
    AS0592 & $-1.98$ & 0.88 & 3.1 & 0.0 & 0.61 & 13.5 & 2.7 & 2.6\\
    
    AS1063 & $-2.04$ & 1.9 & 2.3 & 0.11 & 0.29 & 17.9 & 2.0 & 1.2\\
    
    1ES 0657--56 & $-2.16$ & 1.41 & 3.5 & 0.0 & 0.70 & 15.8 & 2.8 & 1.64\\
    
    RXCJ0232.2--4420 & $-2.25$ & 1.2 & 2.36 & 0.45 & 0.41 & 14.4 & 1.6 & 1.3\\
    
    RXCJ2031.8--4037 & $-2.02$ & 1.1 & 2.4 & 0.12 & 0.30 & 9.4 & 1.3 & 0.75\\
    
    \hline
    \end{tabular}
\end{table*}

\section{Temperatures from X-ray spectral fit (T$_{\rm X-ray}$) vs. our measurements of T$_{\rm X-ray+SZ}$}\label{sec:tsz_txray}
In this section, we compare the global projected temperatures of the galaxy clusters in our sample, measured using Chandra and XMM-Newton spectral fitting (\(T_{\rm X-ray}\)), with our global temperature measurements obtained from X-ray + SZ joint fitting. 
First, we project the best-fit 3D temperature profiles (Figure \ref{fig:pressure_density}) onto the sky plane, following \citet{2021MNRAS.501.3767S} and assuming a spherical distribution,
\begin{equation}
    T_{\rm X-ray+SZ} = 
    \frac{\int{{n}_{e}^2\ 
    {T}_{e}^{\frac{1}{4}}\
    {dV}}}{\int{{n}_{e}^2\ 
    {T}_{e}^{-\frac{3}{4}}\
    {dV}}},
\end{equation}
where $T_e$ and $n_e$ are the
best-fit 3D temperature and 
density profiles.
Finally we measure average 
projected temperature of the 
cluster within 
0.15R$_{500}$--R$_{200}$.
Figure \ref{fig:Tx_vs_Tsz} compares
the resulting average
projected 
temperatures with the global
temperatures
measured using spectral fit from
Chandra (left) and XMM-Newton (right)
observations. 
Our temperature measurements using
joint X-ray and SZ observations 
more closely align with the 
XMM-Newton
measurements than that of Chandra.

\begin{figure}[h!]
    \centering
    \begin{tabular}{cc}
     \includegraphics[width=0.5\textwidth]{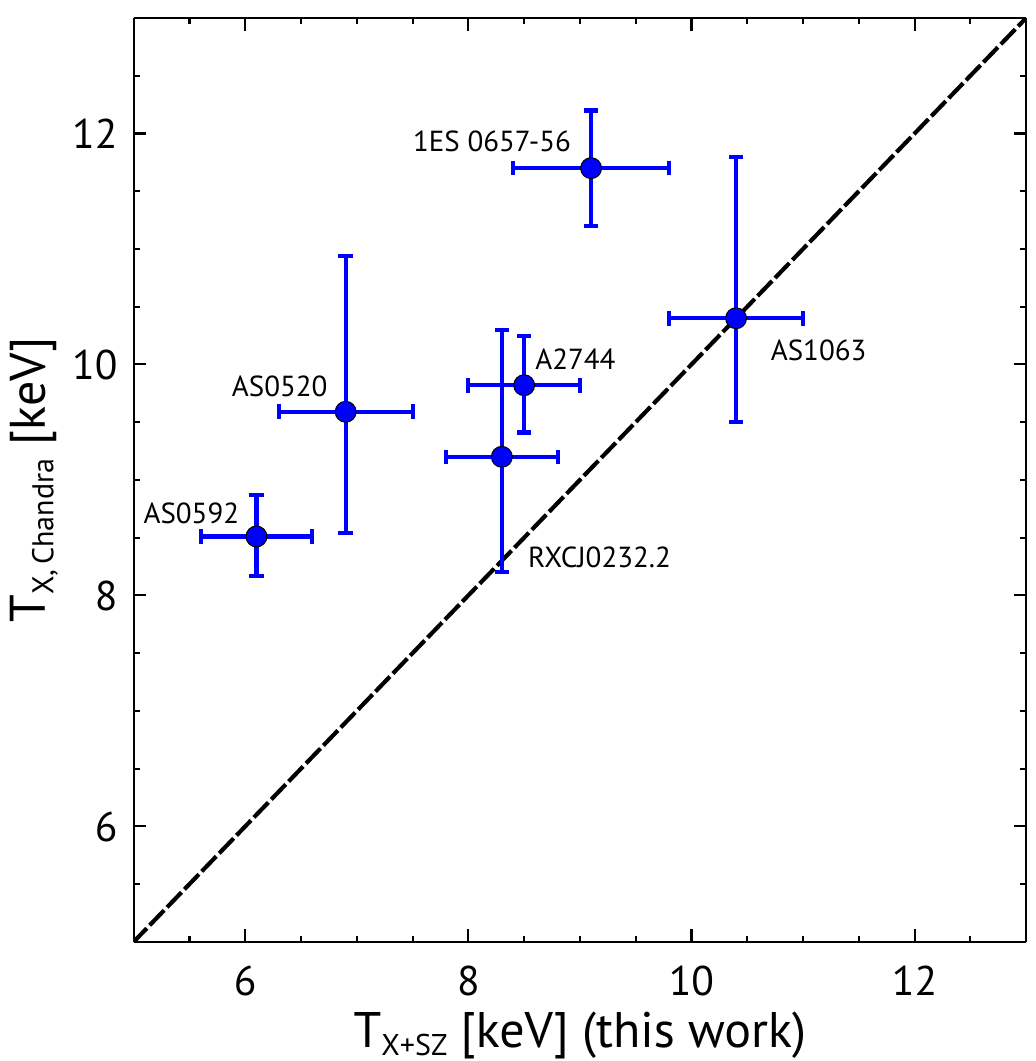} &  \includegraphics[width=0.5\textwidth]{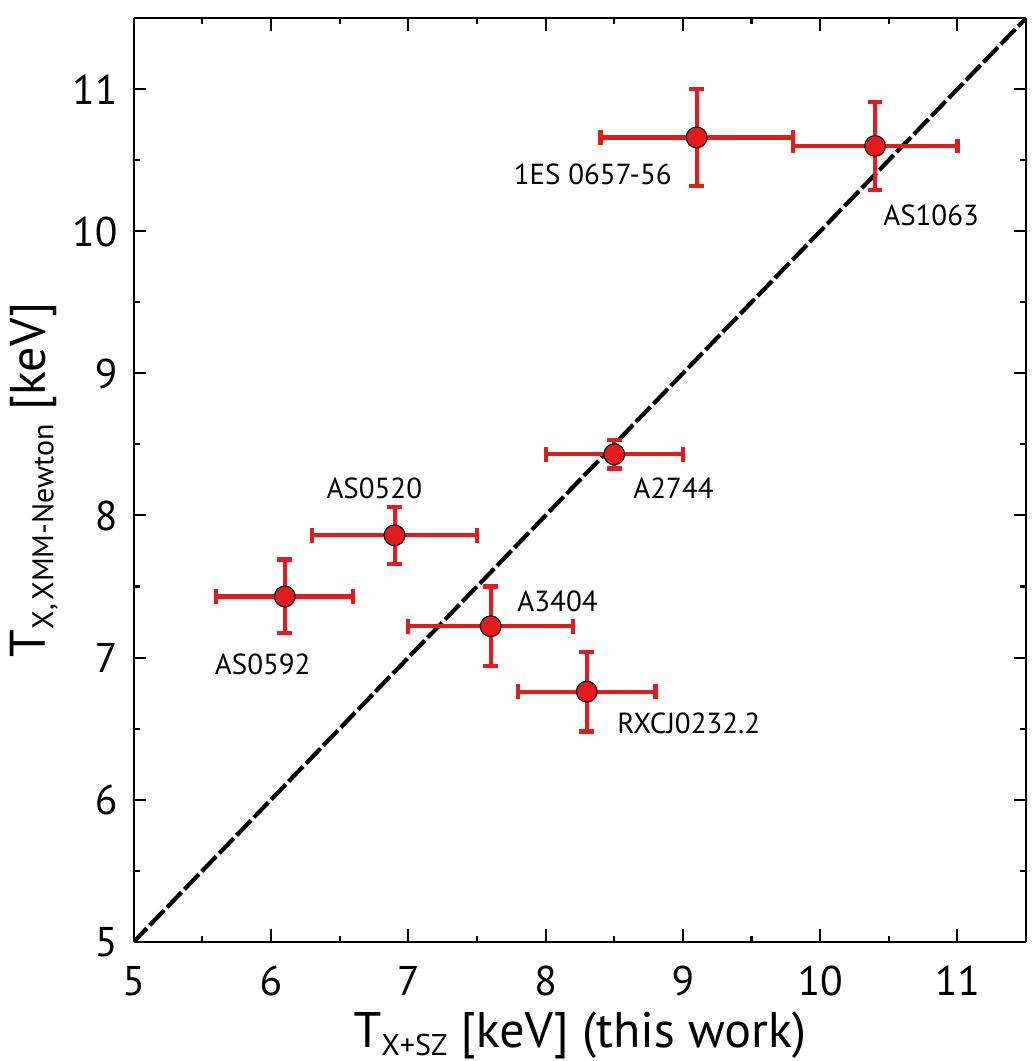} \\
    \end{tabular}
   \caption{Comparison of 
   global projected temperatures 
    within 
    0.15$R_{500} < r < R_{200}$ 
    of clusters
   in our sample measured
    using spectral fitting with
    the projected temperatures 
    within
    same radial range from our 
    X-ray + SZ joint fitting.
    {\it Left}: X-ray temperatures
    are measured using 
    Chandra observations and are
    taken from \citet{2012AASP....2..188B} (A2744), \citet{2010ApJ...723.1523M} (AS0520, AS0592), and 
    \citet{2008ApJS..174..117M} (AS1063, 1ES 0657-56, RXCJ2032.2--4420).
    {\it Right}: X-ray temperatures
    are measured using 
    XMM-Newton observations and are
    taken from \citet{2020ApJ...892..102L}.
    }
   \label{fig:Tx_vs_Tsz}
\end{figure}
}







\label{lastpage}

\end{document}